\documentclass[apj]{emulateapj}

\usepackage[utf8]{inputenc}
\usepackage{graphicx}
\usepackage{captcont}
\usepackage[english]{babel}
\usepackage{amssymb}
\usepackage{float}
\usepackage{multirow}
\usepackage{color}
\usepackage{threeparttable}
\usepackage{color}
\usepackage{arydshln}
\usepackage{grffile}

\definecolor{ashgrey}{rgb}{0.7, 0.75, 0.71}
\definecolor{darkgray}{rgb}{0.66, 0.66, 0.66}
\definecolor{trolleygrey}{rgb}{0.5, 0.5, 0.5}
 	\definecolor{gray}{rgb}{0.5, 0.5, 0.5}
\definecolor{davysgrey}{rgb}{0.33, 0.33, 0.33}
 	
\usepackage[normalem]{ulem}

\slugcomment{}

\shorttitle{Formation of Terrestrial Planets}
\shortauthors{Izidoro et al.}

\begin{document}

\title{Terrestrial planet formation constrained by Mars and the structure of the asteroid belt}

\author{André Izidoro}
\affil{Université de Bordeaux, Laboratoire d’Astrophysique de Bordeaux, UMR 5804, F-33270, Floirac, France. \\ Capes Foundation, Ministry of Education of Brazil, Brasília/DF 70040-020, Brazil. \\University of Nice-Sophia Antipolis, CNRS, Observatoire de la Côte d’Azur, Laboratoire Lagrange, BP 4229,
06304 Nice Cedex 4, France.}
\email{izidoro.costa@gmail.com}

\author{Sean N. Raymond}
\affil{CNRS, Laboratoire d'Astrophysique de Bordeaux, UMR 5804, F-33270 Floirac, France \\
Université de Bordeaux, Laboratoire d'Astrophysique de Bordeaux, UMR 5804, F-33270 Floirac, France}

\author{Alessandro Morbidelli }
\affil{University of Nice-Sophia Antipolis, CNRS, Observatoire de la Côte d’Azur, Laboratoire Lagrange, BP 4229, 06304 Nice Cedex 4, France.}

\author{Othon C. Winter}
\affil{UNESP, Univ. Estadual Paulista - Grupo de Dinâmica Orbital \& Planetologia, Guaratinguetá, CEP 12.516-410, São Paulo, Brazil}

\begin{abstract}

Reproducing the large Earth/Mars mass ratio requires a strong mass depletion in solids within the protoplanetary disk between 1 and 3 AU.  The Grand Tack model invokes a specific migration history of the giant planets to remove most of the mass initially beyond 1 AU and to dynamically excite the asteroid belt.  However, one could also invoke a steep density gradient created by  inward drift and pile-up of small particles induced by gas-drag, as has been proposed to explain the formation of close-in super Earths. Here we show that the asteroid belt's orbital excitation provides a crucial constraint against this scenario for the Solar System. We performed a series of simulations of terrestrial planet formation and asteroid belt evolution starting from disks of planetesimals and planetary embryos with various radial density gradients and including Jupiter and Saturn on nearly circular and coplanar orbits.  Disks with shallow density gradients reproduce the dynamical excitation of the asteroid belt by gravitational self-stirring but form Mars analogs significantly more massive than the real planet. In contrast, a disk with a surface density gradient proportional to $\rm {\it r}^{-5.5}$ reproduces the Earth/Mars mass ratio but leaves the asteroid belt in a dynamical state that is far colder than the real belt. We conclude that no disk profile can simultaneously explain the structure of the terrestrial planets and asteroid belt. The asteroid belt must have been depleted and dynamically excited by a different mechanism such as, for instance, in the Grand Tack scenario.

\end{abstract}


\section{Introduction}

One goal of the field of planet formation is to reproduce the Solar System using numerical simulations (for recent reviews, see Morbidelli et al., 2012 and Raymond et al., 2014).  Most studies have not achieved this goal.  A number of fundamental properties of the Solar System are difficult to replicate.  In this paper we focus on two of these key constraints: Mars' small mass and the structure of the asteroid belt (orbital distribution, low total mass).

The classical scenario of terrestrial planet formation suffers from the so-called ``Mars problem'' (e.g. Chambers, 2014).  Assuming that planets accrete from a disk of rocky planetesimals and planetary embryos that stretches continuously from $\sim$0.3-0.7 AU to about 4-5 AU, simulations consistently reproduce the masses and orbits of Venus and Earth (Wetherill 1978, 1986, Chambers \& Wetherill 1998; Agnor et al., 1999; Chambers, 2001, Raymond et al., 2004, 2006, 2007, 2009; O'Brien et al., 2006; Morishima et al., 2008, 2010; Izidoro et al., 2013; Lykawaka et al., 2014; Fischer \& Ciesla 2014).  However, planets in Mars' vicinity are far larger than the actual planet.  Several solutions to this problem have been proposed, each invoking a depletion of solids in the Mars region linked to either the properties of the protoplanetary disk (Jin et al., 2008; Hansen 2009; Izidoro et al., 2014), perturbations from the eccentric giant planets (Raymond et al., 2009; Morishima et al., 2010), or a combination of both (Nagasawa et al., 2005; Thommes et al., 2008).  Most of these models are either not self-consistent or are simply inconsistent with our current understanding of the global evolution of the Solar System (for a discussion see Morbidelli et al., 2012).  

To date the most successful model -- called the {\em Grand Tack} -- invokes a truncation of the disk of terrestrial building blocks during the inward-then-outward migration of Jupiter in the gaseous protoplanetary disk (Walsh et al., 2011; Pierens \& Raymond 2011; O'Brien et al., 2014; Jacobson \& Morbidelli 2014; Raymond \& Morbidelli 2014).  If the terrestrial planets formed from a truncated disk, then Mars' small mass is the result of an ``edge'' effect, as this planet was scattered out of the dense annulus that formed Earth and Venus (Wetherill, 1978; Hansen 2009; and  Morishima et al., 2008).  

A second model constraint comes from the asteroid belt, specifically the distribution of orbital inclinations.  The present-day main belt has a broad inclination distribution, spanning continuously from $i~=~0^\circ$ to $i~=~20^\circ$.  Figure 1 shows the orbital distribution of the real population of asteroids with diameter larger than about 50 km. The inclinations of most of these objects are far larger than would be expected from formation in a dissipative protoplanetary disk (Lecar \& Franklin 1973).

 At least four models have been proposed for explaining the belt's depletion and excitation (for a detailed discussion see a review by Morbidelli et al., 2015): sweeping secular resonances\footnote{ A secular resonance occurs when the period of the nodal (longitude of the ascending node) or apsidal (longitude of perihelion) precession of the orbit of a small body becomes equal to those of one the giant planets. Equal apsidal frequencies results  in a pumping of the orbital eccentricity of the smaller object, while that a match between nodal frequencies tend to increase the orbital inclination of the smaller body.} driven by the depletion of the nebula (Lecar \& Franklin 1973, 1997), scattering of Earth masses objects by the forming Jupiter (Safronov, 1979; Ip, 1987; Petit et al., 1999), scattering by protoplanetary embryos embedded in the belt (Wetherill 1992; Chambers \& Wetherill 2001; Petit et al., 2001) and the inward scattering of planetesimals during the outward migration of Jupiter and Saturn (i.e., the Grand Tack model: Walsh et al., 2011, 2012).  Two of these models have been discarded because they do not reproduce the observed belt structure: sweeping of secular resonances and planets scattered by Jupiter (see O'Brien et al., 2007 and Morbidelli et al., 2015). For example, O'Brien et al., (2007) showed that sweeping secular resonances  are incapable of giving the observed orbital excitation of the asteroid belt unless the timescale for nebular gas depletion is much longer ($\sim$20 Myr) than the values derived from current observations (1-10 Myr).

Within the classical scenario of terrestrial planet formation, the most successful model to excite and deplete the belt invokes the existence of planetary embryos in the primitive asteroid belt (Chambers \& Wetherill 2001; Petit et al., 2001; O'Brien et al., 2007).  These putative planetary embryos depleted the main belt and excited the eccentricities and inclinations of the orbits of surviving asteroids. In this scenario, the asteroid belt is initially massive (roughly 1-2 Earth masses, as numerous massive planetary embryos are needed to excite the belt), while it is mass-depleted in the end. The initial presence of a significant amount of mass in the asteroid belt, though, has the drawback of producing a too massive planet at the location of Mars. In addition, all planetary embryos need to be removed from the belt on a 100 Myr timescale; if planetary embryos are removed on a significantly longer timescale they carve a ``hole'' in the observed asteroid distribution (Petit et al., 1999; Raymond et al., 2009).

\begin{figure}
\centering
\includegraphics[scale=.7]{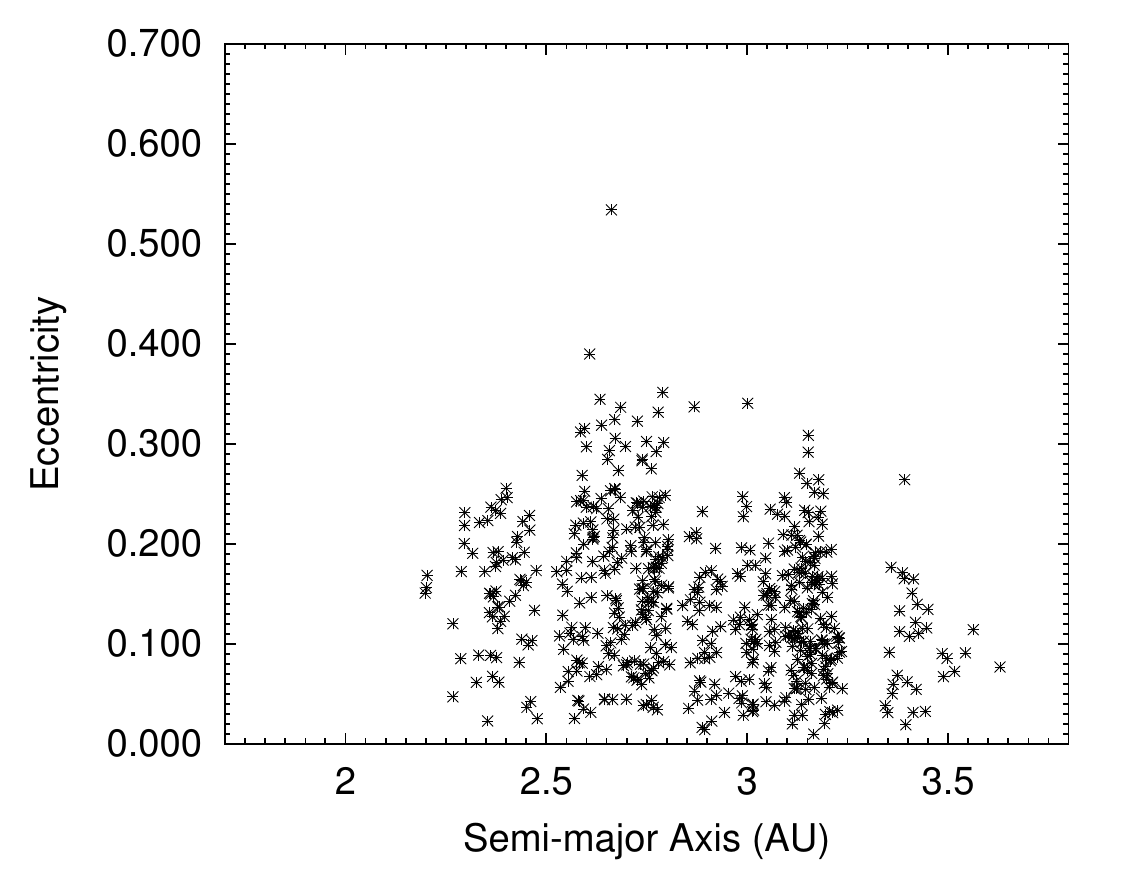}
\includegraphics[scale=.7]{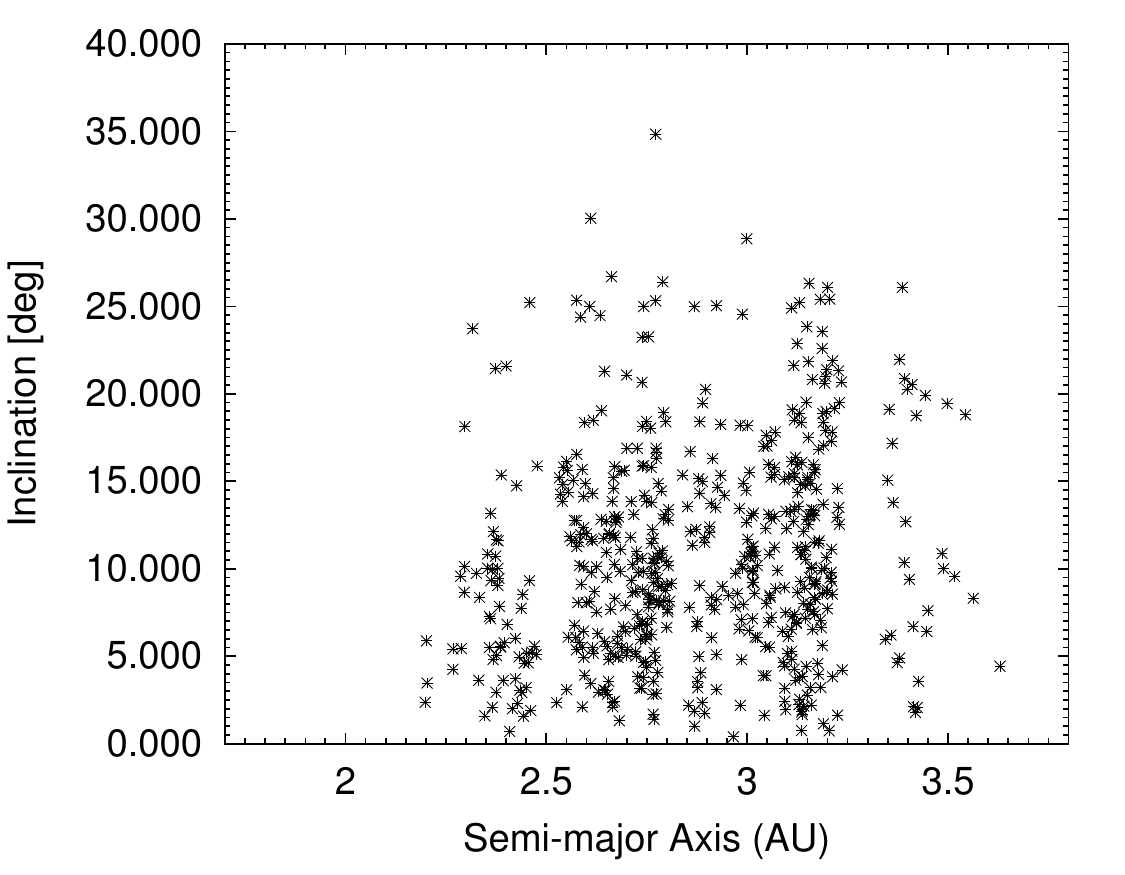}
\caption{Orbital distribution of the real population of asteroids with absolute magnitude $H<9.7$, which corresponds to diameter of about 50 km. The upper plot show semi-major axis vs. eccentricity. The lower plot shows semimajor axis vs inclination.}
\end{figure}

Unlike the classical scenario, the orbital excitation and mass depletion of the asteroid belt in the Grand-Tack model are both essentially created  during the inward-then-outward migration of Jupiter and Saturn (Walsh et al., 2011; Jacobson \& Walsh, 2015). In this model, the giant planets crossed the asteroid belt twice.  First, during their inward migration stage, the gas giants compress the distribution of planetary embryos and planetesimals inside Jupiter's orbit into a narrow disk around 1 AU. A fraction of these objects is also scattered outwards. Second, during the outward migration phase, the giant planets scatter inwards a fraction of planetesimals beyond 2-3 AU enough to repopulate the asteroid belt region, with a dynamically excited population of small bodies carrying altogether a small total mass.

 An alternative to the Grand Tack scenario to produce the confined disk and a mass deficient asteroid belt could be invoked that a lot of solid material drifted to within 1 AU by gas drag, leaving the region beyond 1 AU substantially depleted in mass. This idea is very appealing in a broad context of planet formation. This is because it is often invoked to produce a large pile-up of mass in the inner disk to explain the formation of close-in super-Earths (e.g. Chatterjee \& Tan 2014; Boley \& Ford, 2014). Moreover, it could also be consistent with modern ideas on planetesimal formation and planetary growth  based on the drift and accretion of pebbles (Lambrechts \& Johansen, 2014; Lambrechts et al., 2014; Johansen et al., 2015). Particles drifting toward the star can produce in principle disks of solids of any radial gradient in the resulting mass distribution. Therefore, the goal of this paper is to test whether any of these gradients  could explain at the same time the small mass of Mars and the properties of the asteroid belt (mass deficit and inclination excitation). In other words, can we match these constraints {\em without} invoking a dramatic event within the inner Solar System (such as a Grand Tack)?

This paper is laid out as follows. We describe the details of our model in section 2.  In Section 3 we present and analyze our results. In Section 4 we discuss our results and present our conclusions.

\section{The Model and Numerical Simulations}

 The simulations presented in this paper fit in the context of the classical scenario of terrestrial planet formation. We perform simulations starting from disks with a wide range of surface density profiles.  We do not interpret these disk profiles as reflecting the properties of the primordial gaseous disk, as they are not consistent with viscous disk models (e.g., Raymond \& Cossou 2014).  Rather, we assume that the distribution of solids has been sculpted by other processes such as aerodynamic drift (Adachi et al., 1976; Weidenschilling 1980). In an infinite disk, the drift of particles would create a steady flow and no steep radial gradient of mass. But, if Jupiter formed early it could have acted as a barrier to inward drifting pebbles or planetesimals (or even planetary embryos; Izidoro et al., 2015). Thus, if a pressure bump existed in the terrestrial planet formation zone to stop the inward drift (e.g., Haghighipour \& Boss 2003), these two effects together could potentially have led to the creation of a very steep density profile in the terrestrial planet region and throughout the asteroid belt.  

We tested disks with surface density profiles given by $ \Sigma_1 r^{-x}$, where {\it x~}=~2.5, 3.5, 4.5 or 5.5. ${\Sigma_1}$ is the solid  surface density at 1 AU. Our disks extends  from 0.7 to 4 AU. These profiles are much steeper than previous simulations, which were almost always limited to {\it x~}=~1.5 and 1 (except for Raymond et al., 2005; Kokubo et al., 2006; Izidoro et al., 2014). We adjusted $\Sigma_1$ to fix the total mass in the disk between 0.7 and 4 AU at $2.5 M_\oplus$, comparable to the sum of the masses of the terrestrial planets.

The disk is divided into populations of planetesimals (30-40\% by mass) and planetary embryos (60-70\%). Planetary  embryos are assumed to have formed by oligarchic growth and are thus randomly spaced by 5 to 10 mutual Hill radii (Kokubo \& Ida, 1998; 2000). Individual planetesimals were given masses of 0.00075 Earth-masses or smaller. Planetesimals are assumed to interact gravitationally only with the protoplanetary embryos, giant planets and the star, but not with each other.  The masses of the planetary embryos scale as $M\sim r^{3(2-x)/2}\Delta^{3/2}$ (Kokubo \& Ida 2002; Raymond et al., 2005, 2009) where $\Delta$ is the number of mutual Hill radii separating adjacent orbits. This amounts to roughly 80 planetary embryos and 1000 planetesimals.  Figure 2 shows the initial conditions of our simulations.  The initial individual embryo mass at a given orbital distance can vary between different disks by up to a factor of $\sim$10.  Steeper disks (with higher values of $x$) have more massive planetary embryos in the inner parts of the disk, although they never exceed $0.3 M_\oplus$.  Steep disks also have smaller planetary embryos farther out.  In our steepest disks the embryo mass actually drops below the planetesimal mass in the asteroid region (and they become non-self-gravitating bodies).  The lack of gravitational interactions among non-self-gravitating bodies (necessary in order to keep the computational times manageable) has important implications for this study. We will pay special attention to this issue during the analysis and discussion of our results. The initial orbital inclinations of planetesimals and planetary embryos were chosen randomly from the range of $10^{-4}$ to $10^{-3}$ degrees, and their other orbital angles were randomized. Their initial eccentricities are set equal to zero.

Our simulations also included fully-formed Jupiter and Saturn on orbits consistent with the latest version of the Nice Model (Levison et al., 2011). Their initial semi-major axes were 5.4 and 7.3 AU (2/3 mean motion resonance), respectively. Their eccentricities (and inclinations) were initially $\sim 10^{-2}$ (degrees).  For each surface density profile we performed 15 simulations with slightly different randomly generated initial conditions for planetary embryos and planetesimals. Collisions between planetary objects are always treated as inelastic mergers that conserve linear momentum. The simulations were integrated for 700 Myr using the Symba integrator (Duncan et al., 1998) and a timestep of 6 days. Planetary objects that reach heliocentric distances equal to 120 AU are removed from the system. During our simulations, we neglect gas drag and gas-induced migration of the planetary embryos. Namely, we assume that the initial conditions illustrated in Fig. 2 apply at the disappearance of the gas.

\begin{figure*}
\centering
\includegraphics[scale=.65]{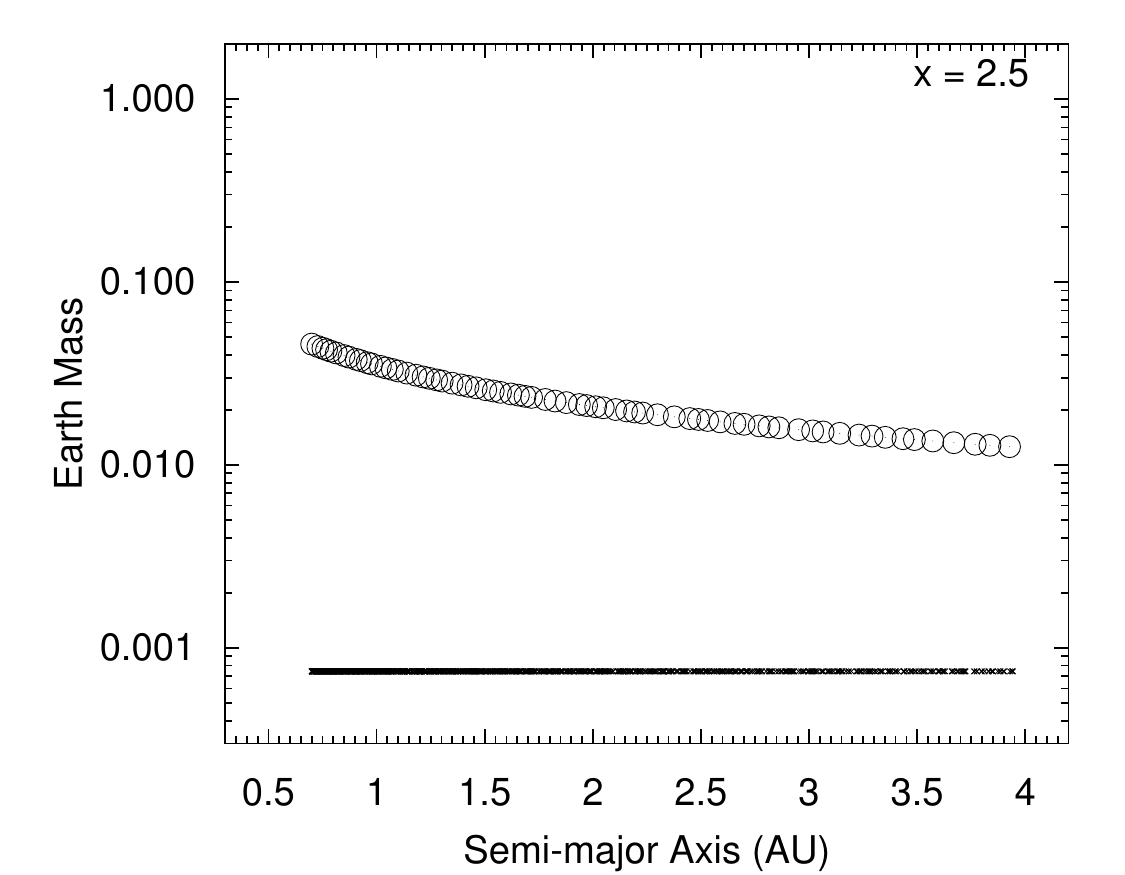}
\includegraphics[scale=.65]{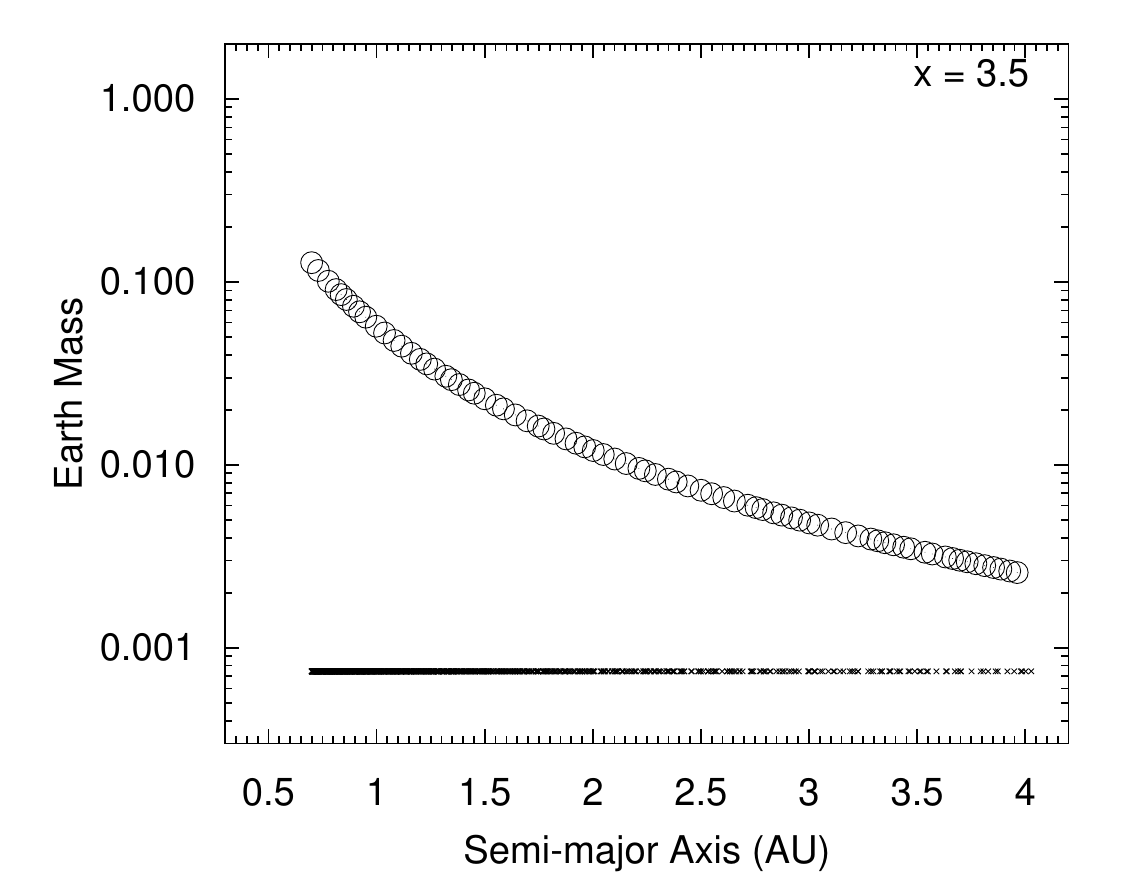}

\includegraphics[scale=.65]{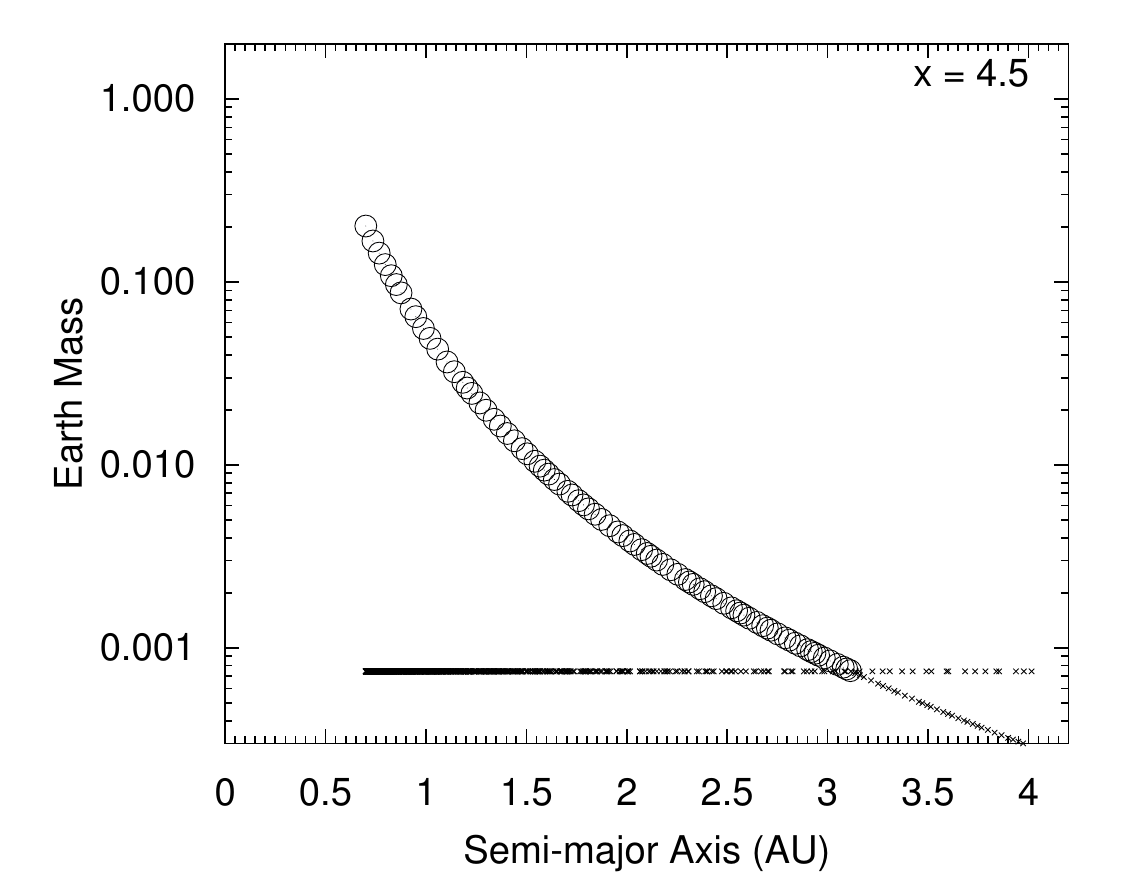}
\includegraphics[scale=.65]{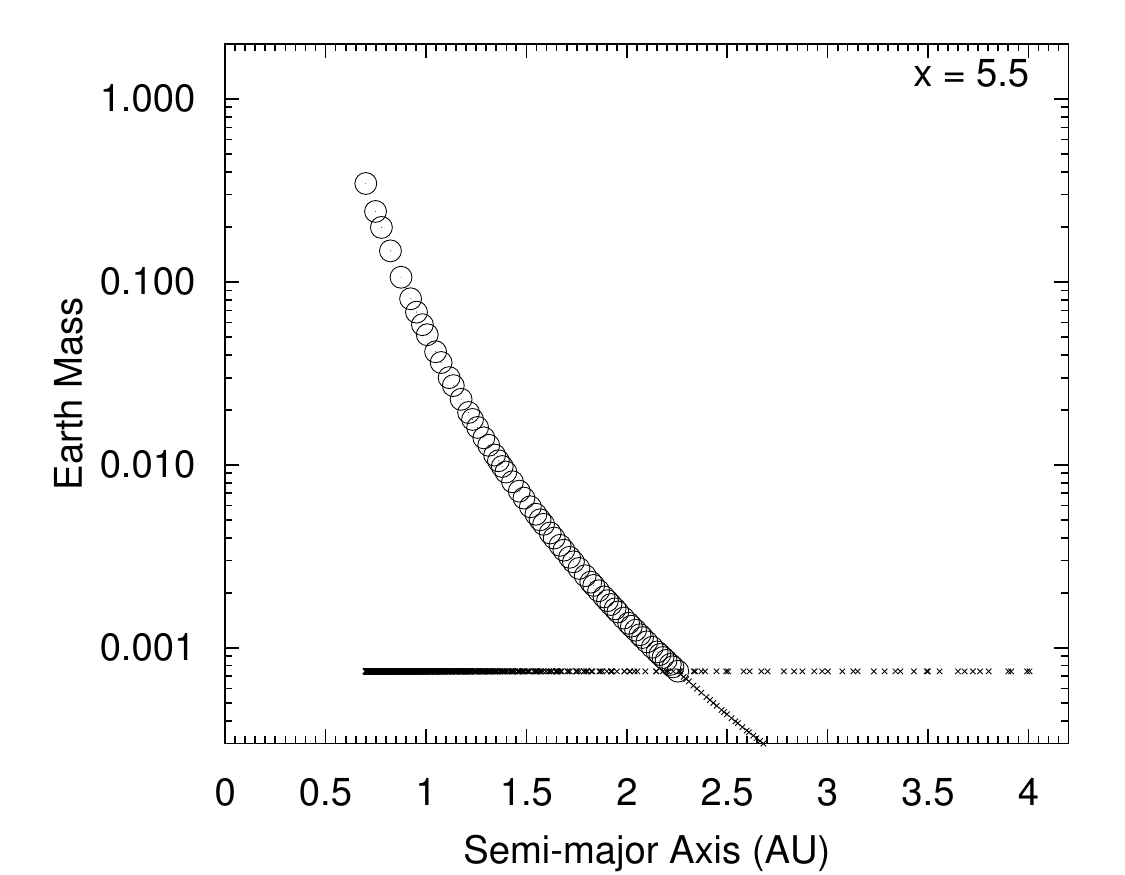}
\caption{Initial conditions of our simulations.  Each panel shows the distribution of planetary embryos and planetesimals generated within a given disk profile defined by the surface density slope $x$. Planetary embryos are marked as open circles and planetesimals are objects with masses smaller than 0.00075 Earth masses.}
\end{figure*}

\section{Results}

Figures 3 and 4 show the evolution of accretion in disks with relatively shallow ({\it x~}=~2.5; Fig 3) and steep ({\it x~}=~5.5; Fig 4) surface density profiles.  A clear difference between these two cases is the accretion timescale of the final planets. In the simulation with {\it x~}=~2.5, the accretion timescale is comparable to that observed in traditional simulations of terrestrial planet formation (over 100 Myr) but in the simulations with {\it x~}=~5.5 growth is much faster (see Raymond et al., 2005, 2007; Kokubo et al., 2006).  This can be understood simply by the fact that steeper disks have more mass in their inner parts, where the accretion timescale -- which depends on the surface density and the dynamical time -- is much shorter. The accretion timescale of terrestrial planets is in this case a few 10 Myr, consistent with some estimates based on radiative chronometers  (Yin et al., 2002; Jacobsen, 2005; see section 3.1.2)

The simulation from Fig 3 formed a planetary system with 3 planets inside 2 AU. The two inner planets at $\sim$0.69 AU and 0.96 AU are reasonable Venus and Earth analogs.  However, the third planet formed around 1.5 AU is about five times more massive than Mars. The Mars problem is pervasive in all simulations with {\it x~}=~2.5~.  This set of simulations is also characterized by the absence of surviving planetesimals in the region of the asteroid belt and, in contrast, the survival of a larger Mars-size planetary embryos in this same region. 

The simulation from Fig. 4 formed a good Mars analog at 1.58 AU with a mass just 30\% larger than Mars'.  As in the previous simulation, a leftover planetary embryo survived beyond the location of Mars, albeit a much smaller one than in the previous case.  Embryos stranded in the asteroid belt could easily be removed during a late instability among the giant planets (i.e., the Nice model).  However, an embryo more massive than $\sim 0.05 M_\oplus$ carves a gap in the asteroid distribution and this gap survives the giant planet instability (O'Brien et al., 2007; Raymond et al., 2009).  

The key result from Figure 4 is that the surviving asteroids have very low eccentricities (and inclinations, as discussed later). Thus, despite forming good Mars analogs, these simulations cannot be considered successful. The surviving planetesimals beyond 2.5 AU show an orbital eccentricity distribution varying from almost zero to 0.1. However, the observed asteroids have an eccentricity distribution that ranges from 0 to 0.3.  The very low asteroidal eccentricities  in the simulation are simply because for such a steep surface density profile there is very little mass in the asteroid belt, and this mass is carried only by planetesimals (only about $0.05 M_\oplus$). No embryo exists beyond $\sim$2.2 AU to stir up planetesimals in this simulation. The inner part of the main belt -- between 2 and 2.5 AU -- does have eccentricities up to 0.2, simply because of their closer proximity to planetary embryos.  One caveat is that our simulated planetesimals do not self-gravitate and therefore cannot self-excite their eccentricities. We will consider self-interacting planetesimals in the asteroid belt in section 4.

\begin{figure*}
\centering
\includegraphics[scale=.65]{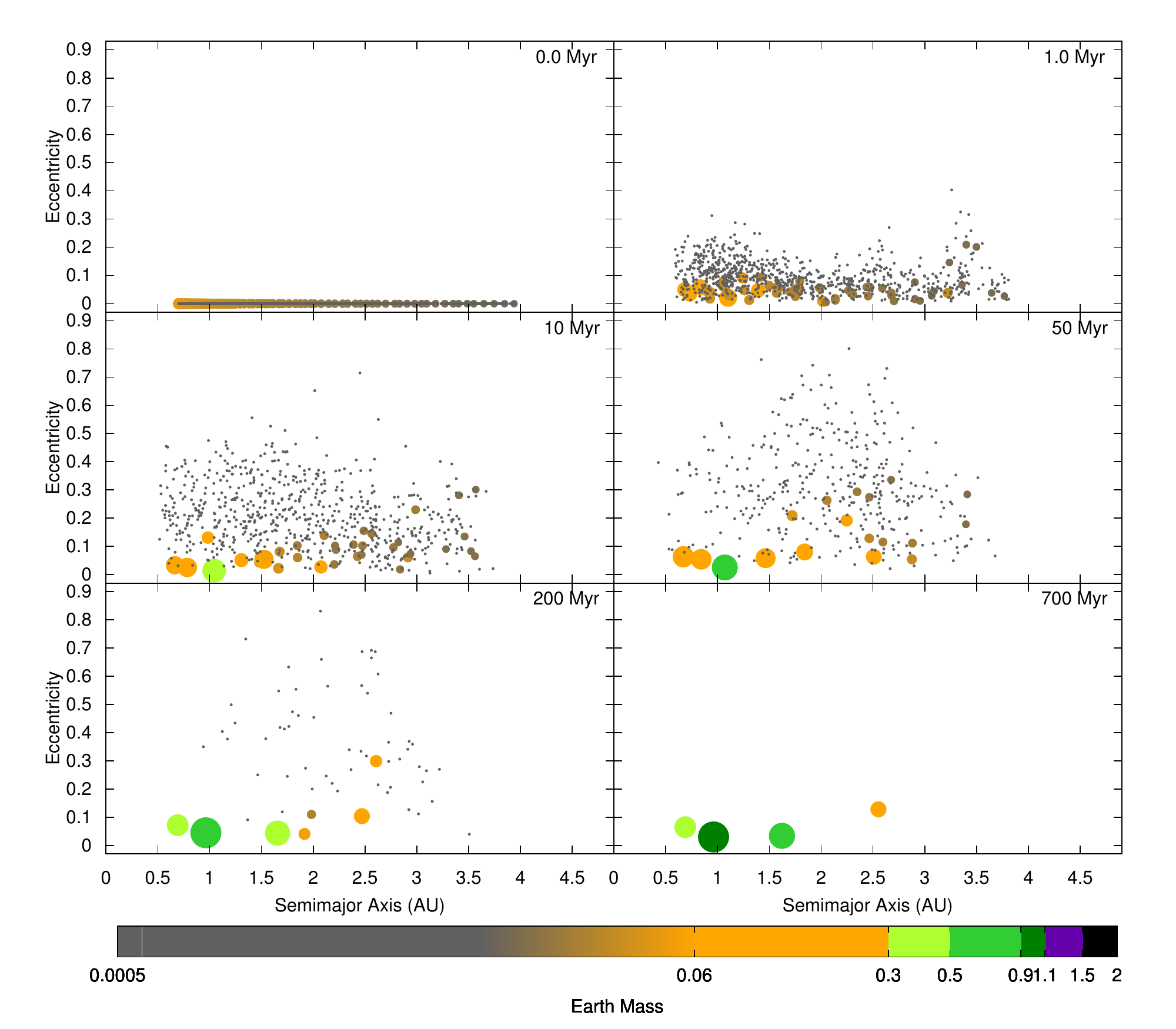}
\caption{Snapshots of the dynamical evolution of one simulation where {\it x~}=~2.5. Jupiter and Saturn are initially as in the Nice Model II. The size of each body corresponds to its relative physical size and is scaled as $M^{1/3}$, where $M$ is the body mass. However, it is not to scale on the {\it x}-axis.  The color-coding gives the range of mass which each body belongs to.}
\end{figure*}

\begin{figure*}
\centering
\includegraphics[scale=.65]{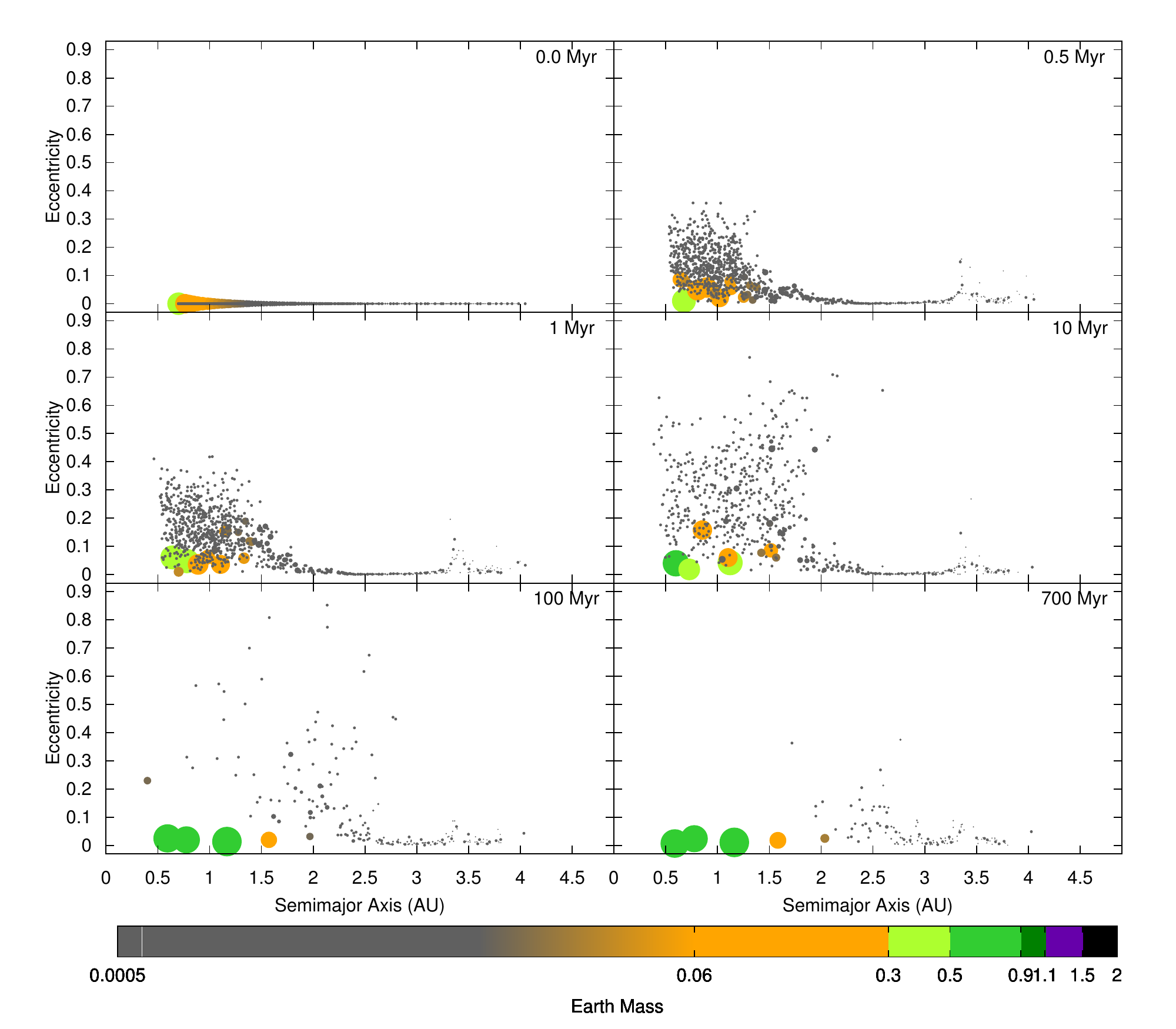}
\caption{Same as Figure 3 but for {\it x~}=~5.5.}
\end{figure*}

\subsection{The final systems}

\begin{figure*}
\centering
\includegraphics[scale=.65]{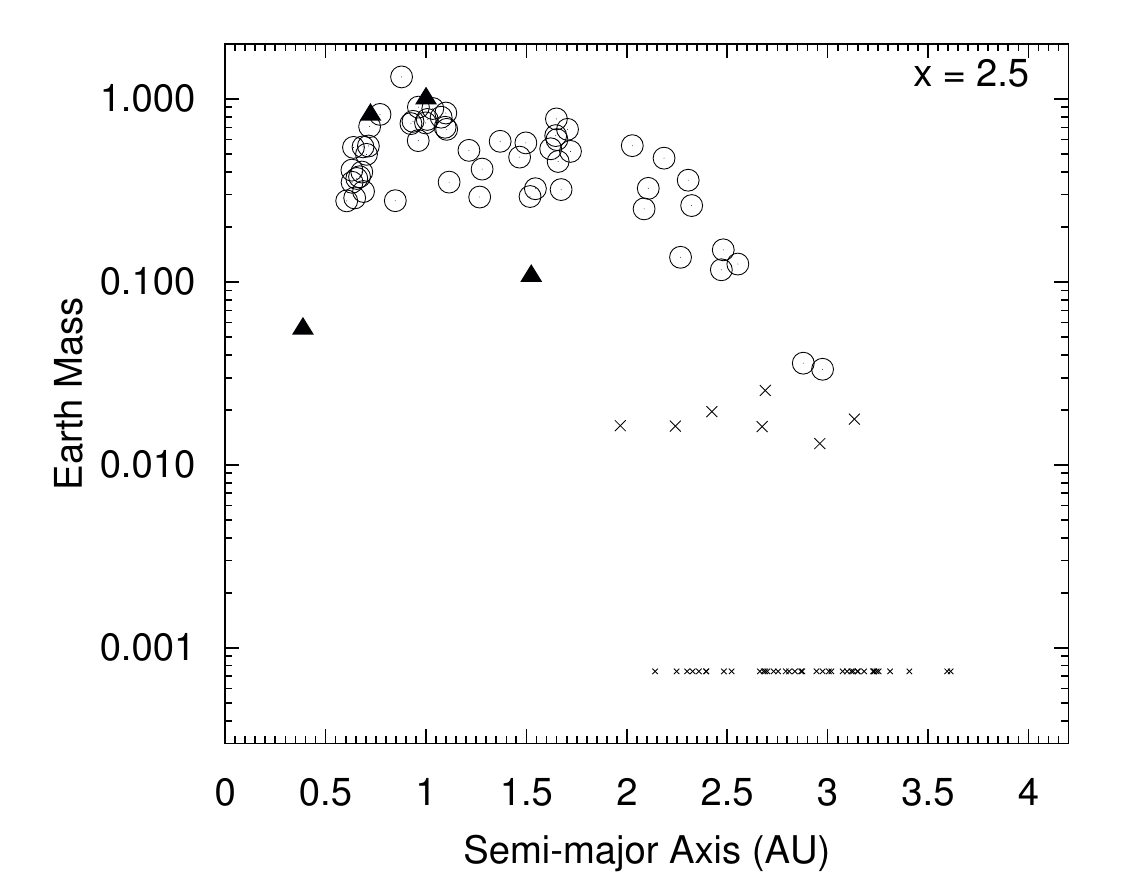}
\includegraphics[scale=.65]{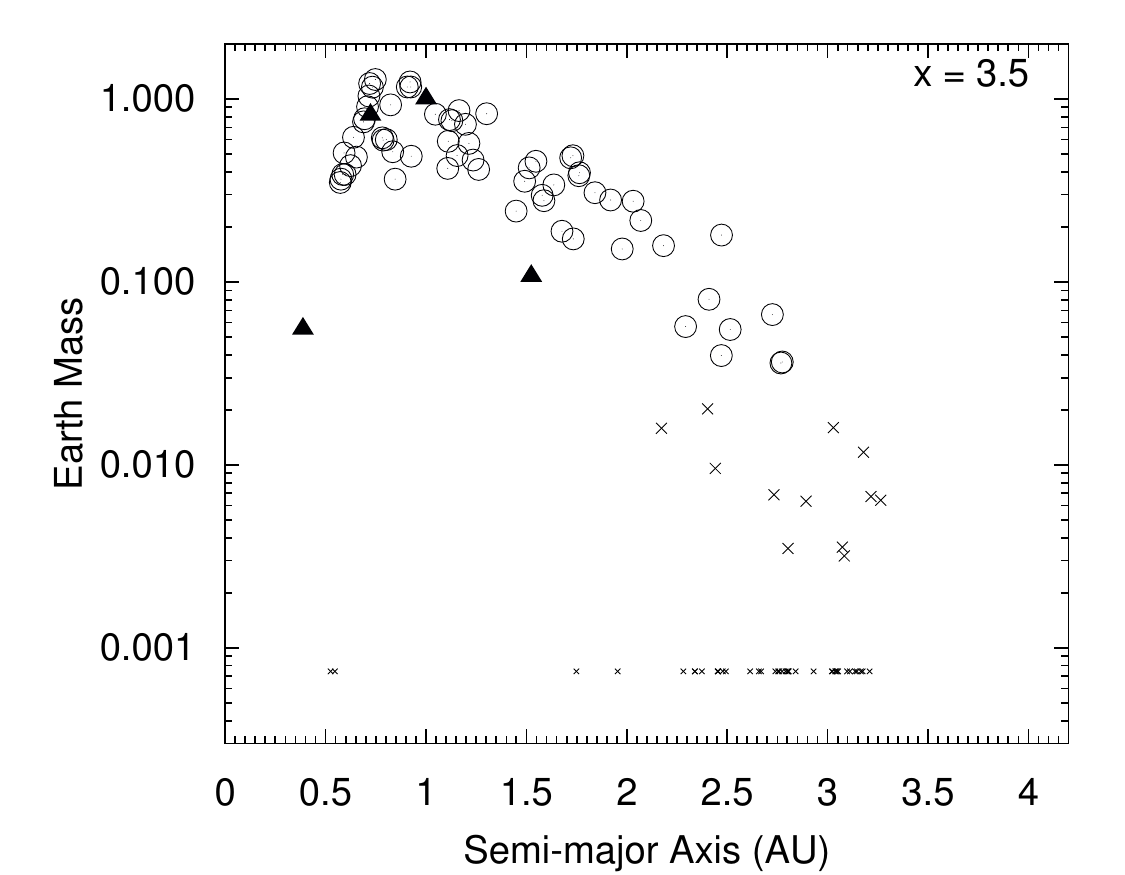}

\includegraphics[scale=.65]{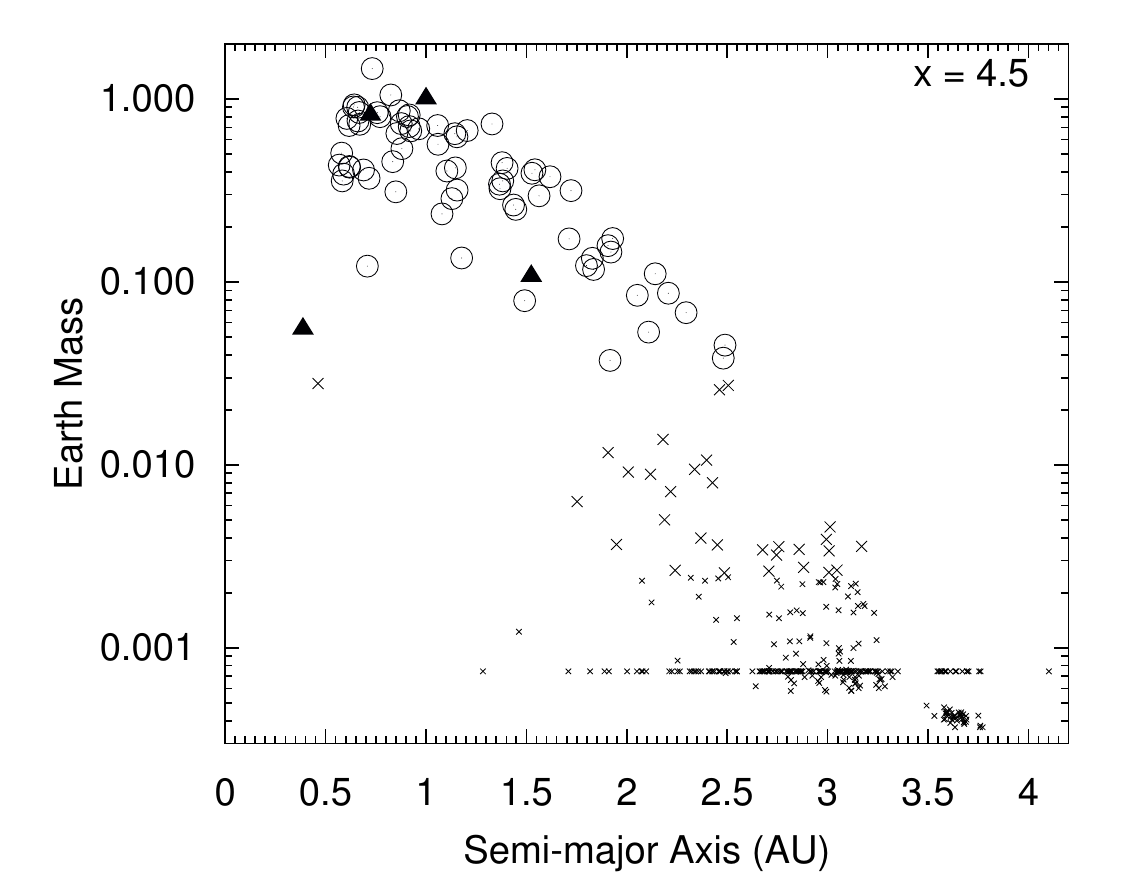}
\includegraphics[scale=.65]{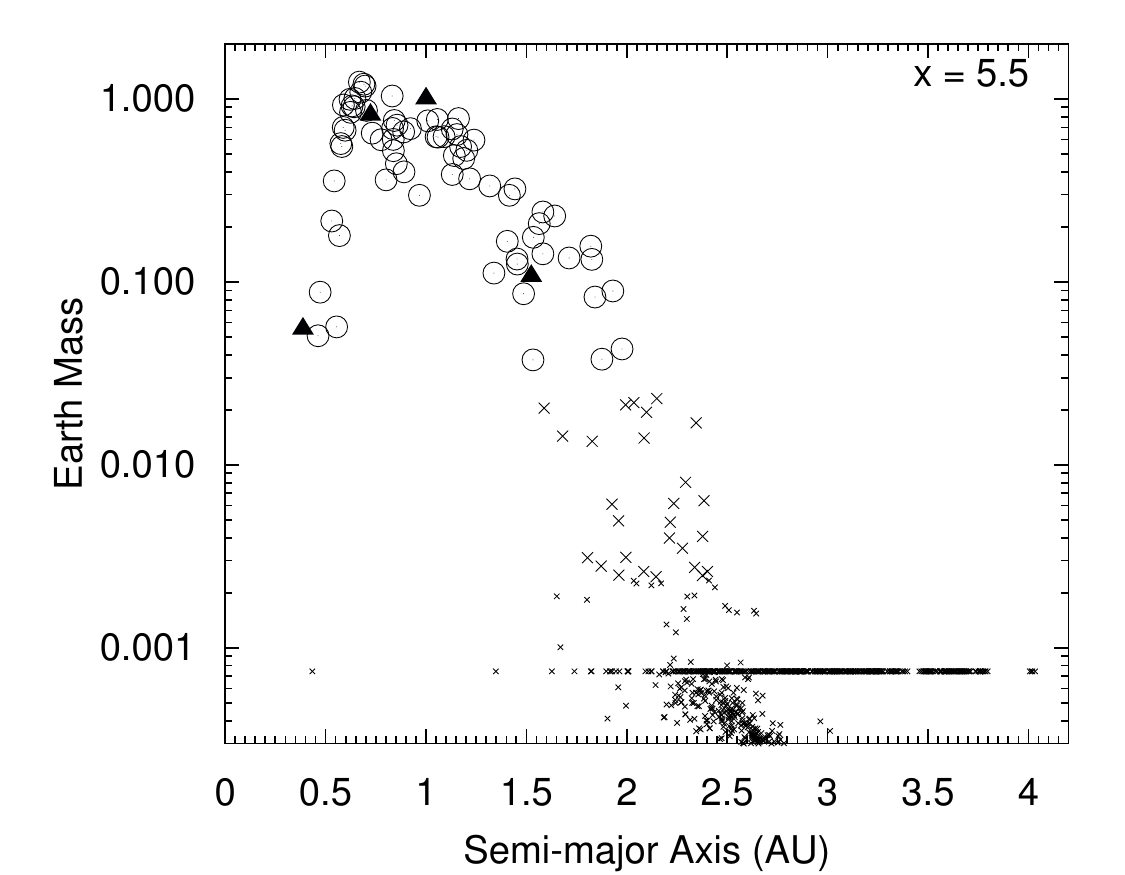}
\caption{Final distribution of the surviving bodies in our simulations after 700 Myr of integration. Each panel shows the final distribution in a diagram semi-major axis versus mass for all simulations considering the same value of {\it x}. The value of {\it x} is indicated on the upper right corner of each panel. Open circles correspond to bodies with masses larger than 0.3 ${\rm M_{\oplus}}$. Smaller bodies are labeled with crosses.  The solid triangles represent the inner planets of the solar system.}
\end{figure*}

\begin{figure*}
\centering
\includegraphics[scale=.65]{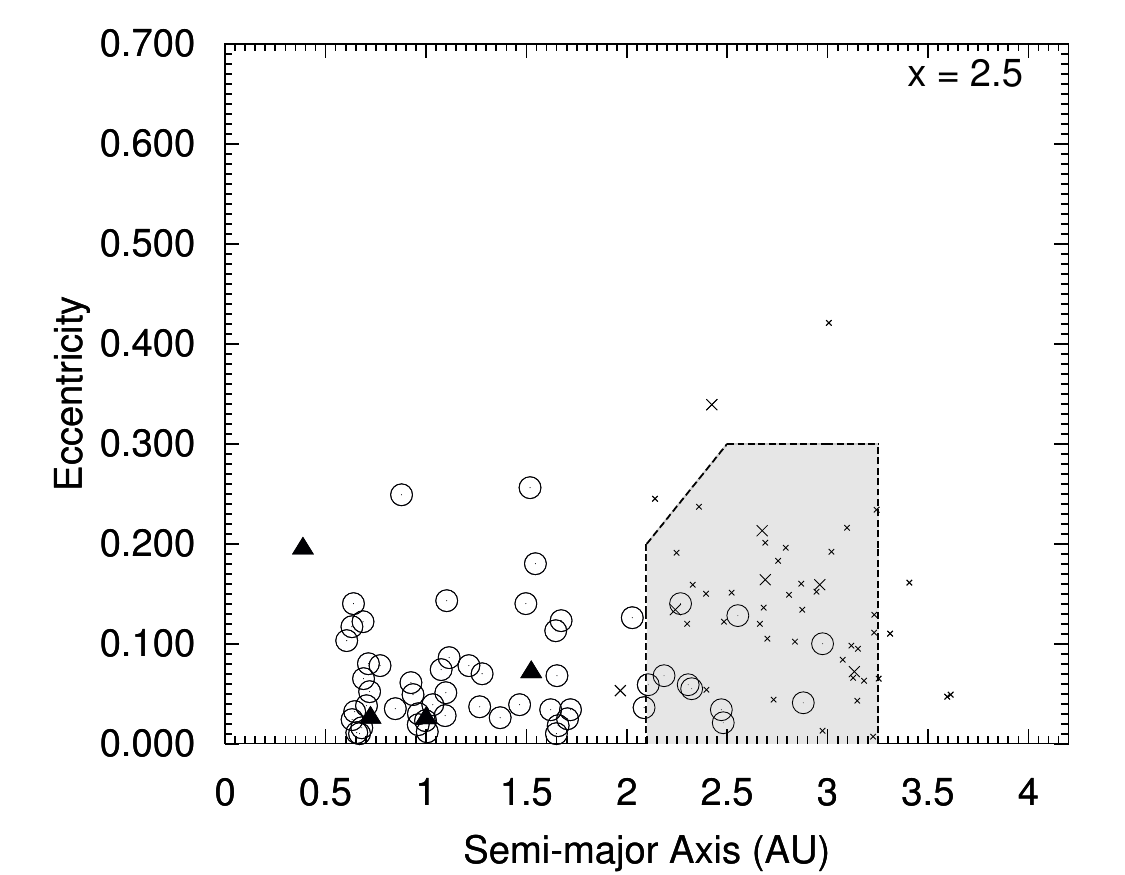}
\includegraphics[scale=.65]{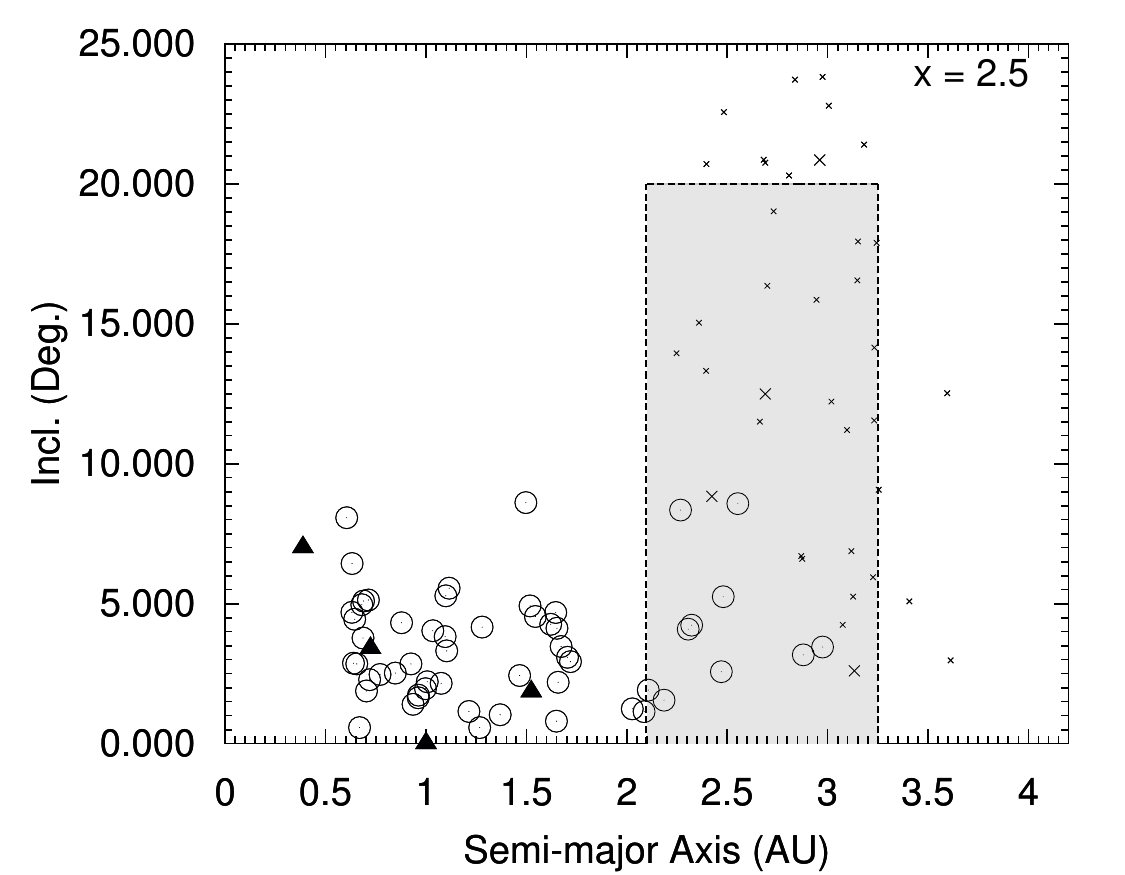}

\includegraphics[scale=.65]{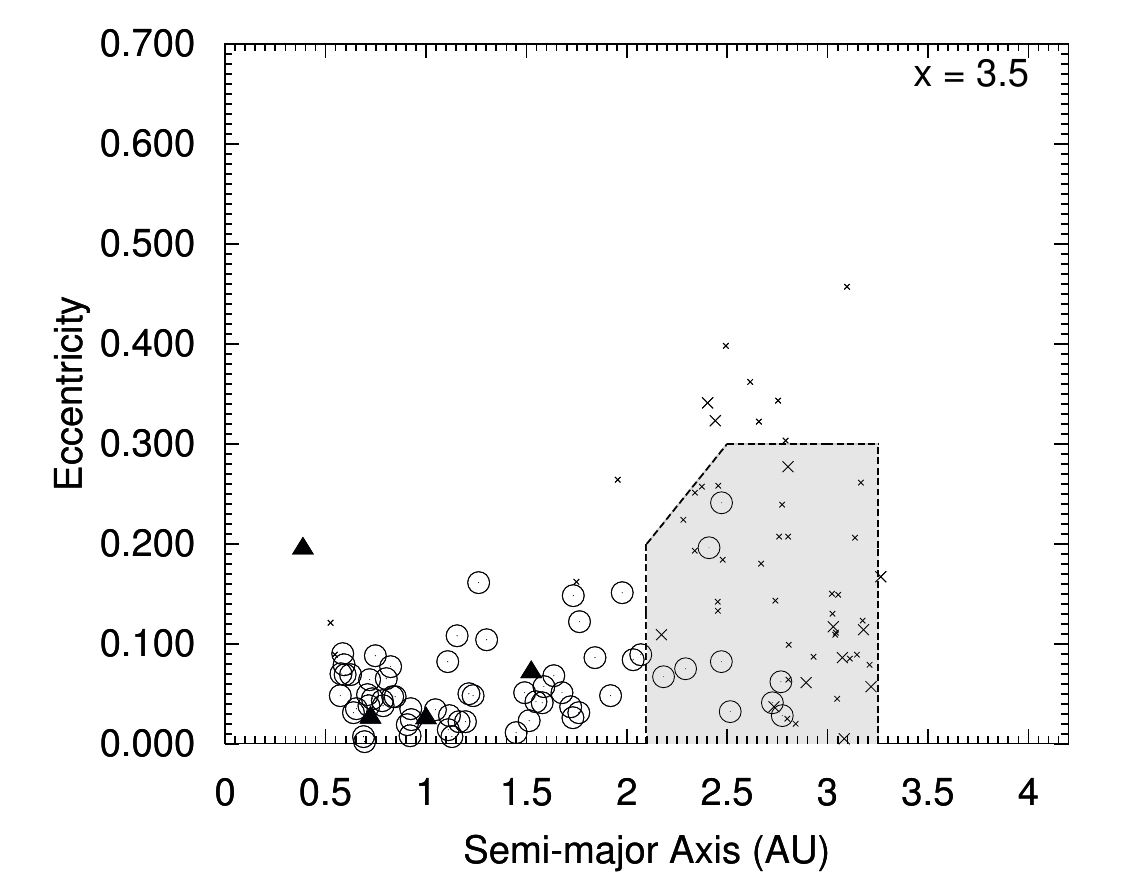}
\includegraphics[scale=.65]{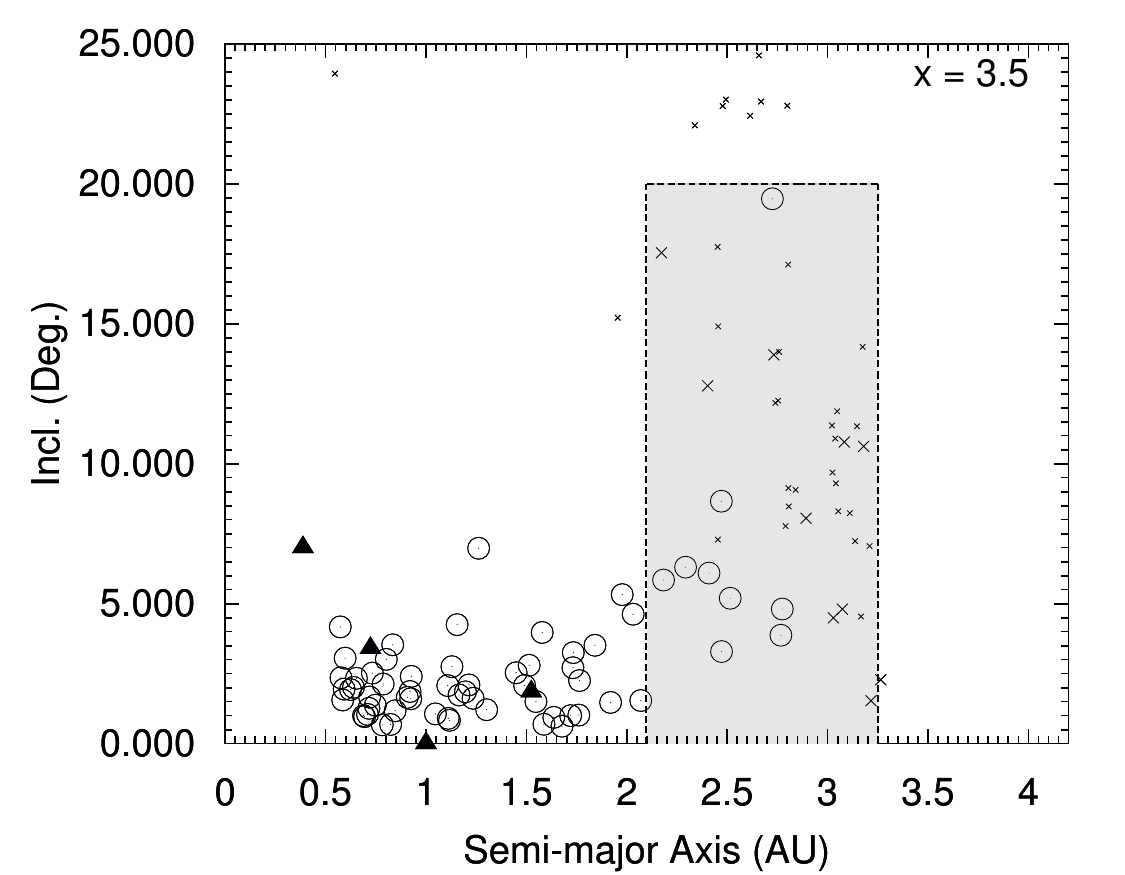}

\includegraphics[scale=.65]{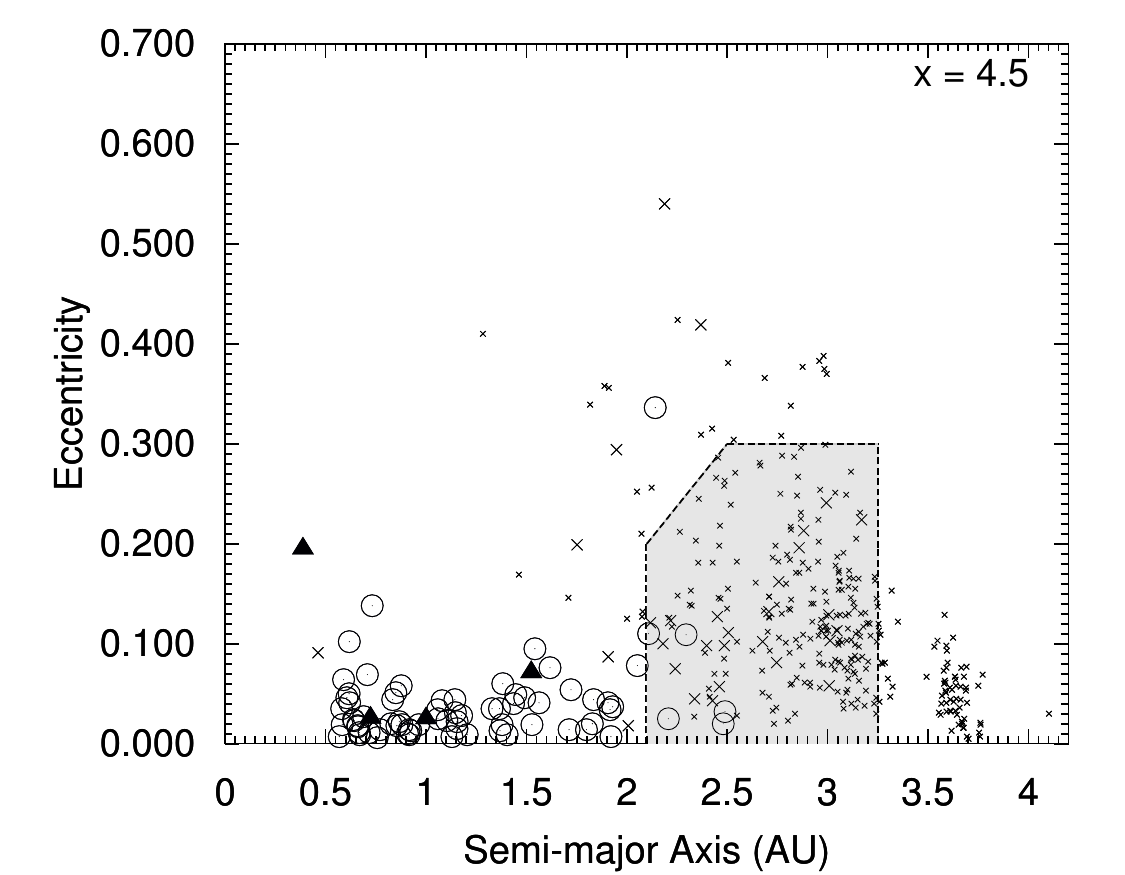}
\includegraphics[scale=.65]{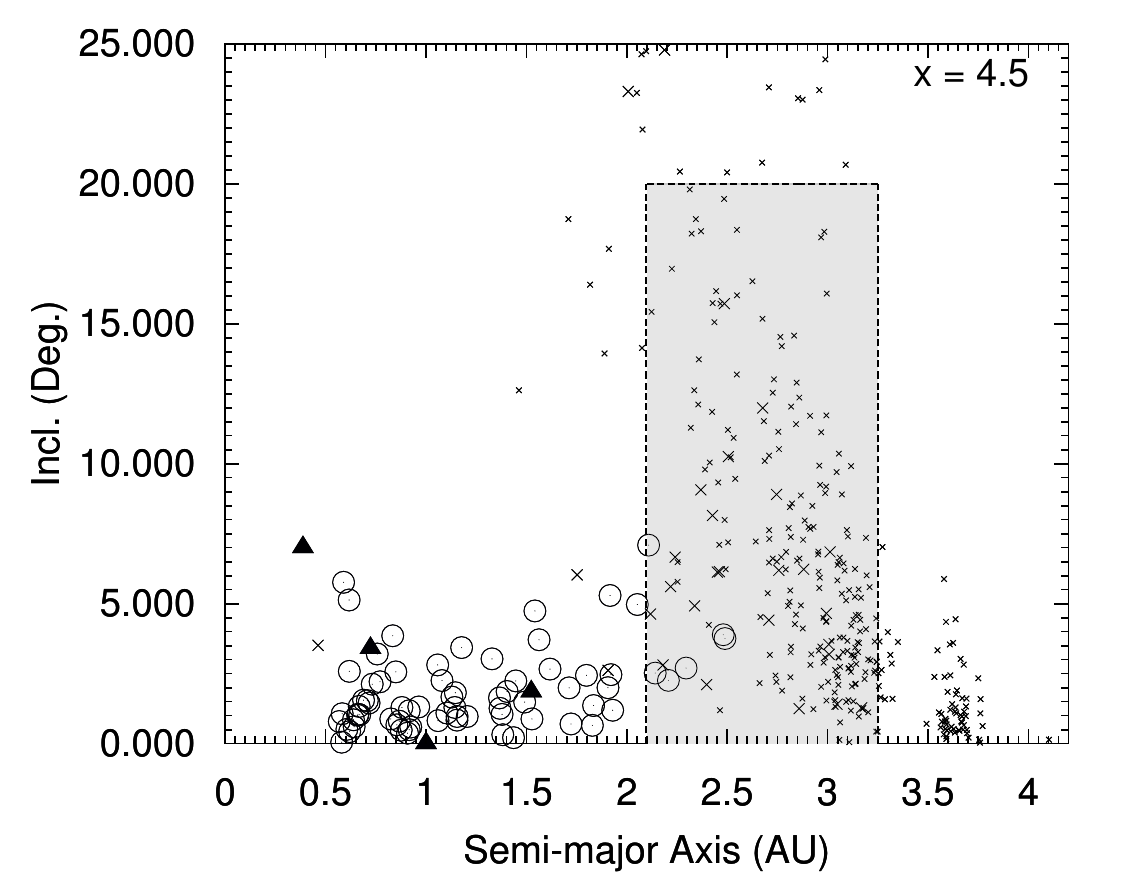}

\includegraphics[scale=.65]{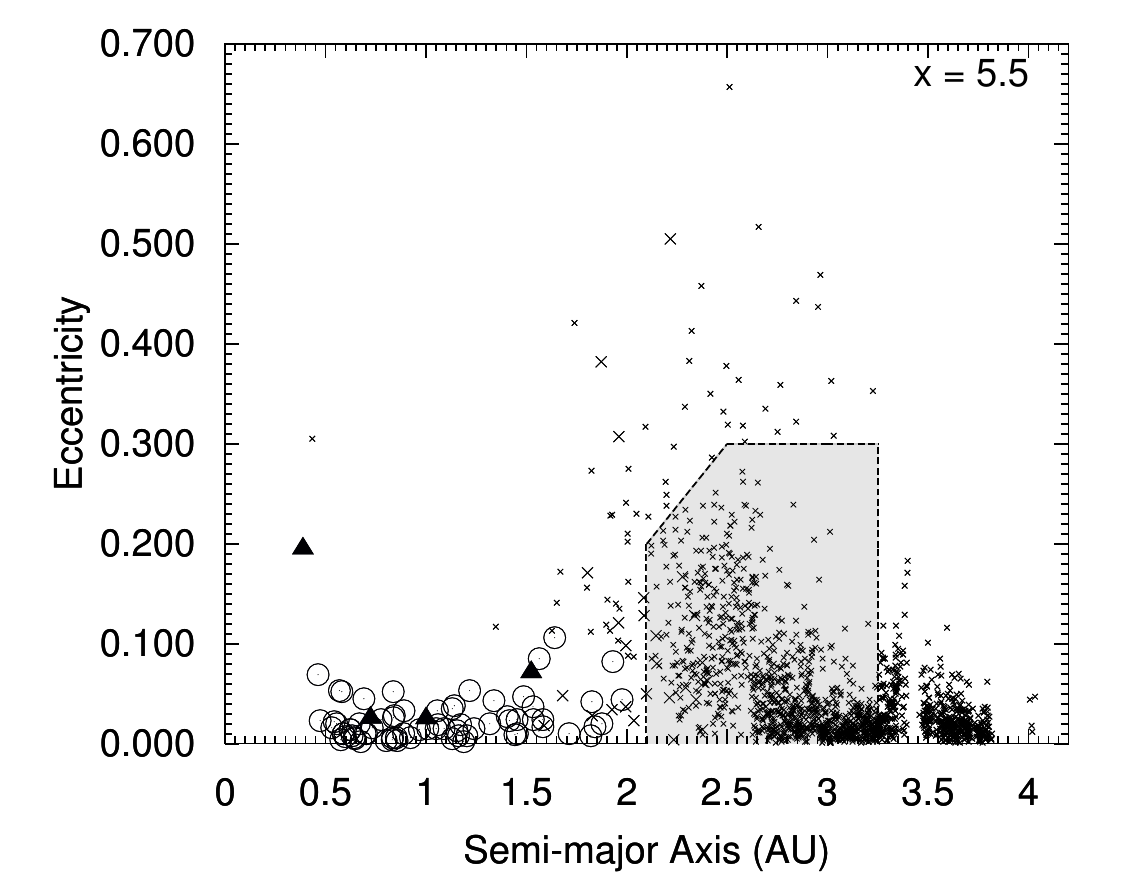}
\includegraphics[scale=.65]{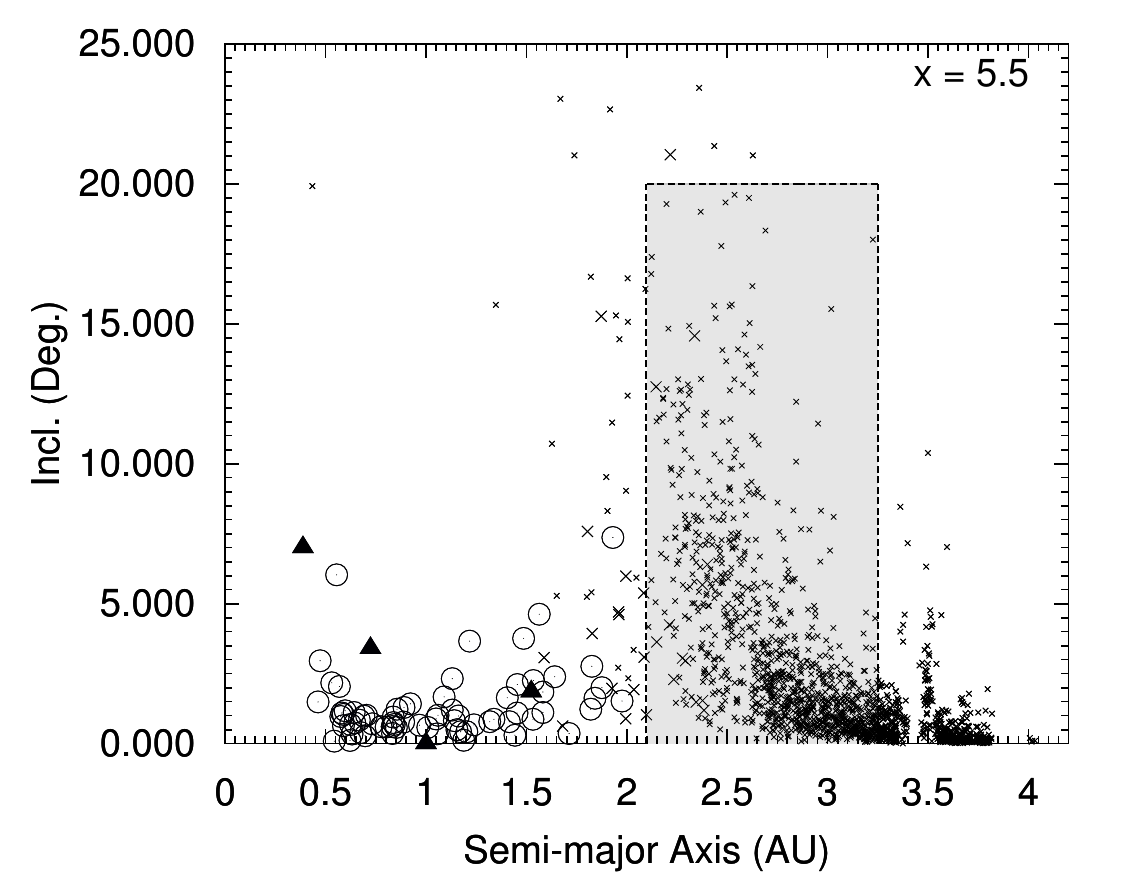}

\caption{Orbital distributions of the surviving bodies in our simulations after 700 Myr of integration. Open
circles correspond to bodies with masses larger than 0.3 ${\rm M_{\oplus}}$.  Smaller bodies are labeled with crosses. The left column shows the orbital eccentricity versus semi-major axis while the right column show orbital inclination versus semi-major axis. The value of $x$ is indicated on the upper right corner  of each panel. The solid triangles represent the inner planets of the solar system. Orbital inclinations are shown relative to a fiducial plane of reference.}
\end{figure*}

Figures 5 and 6 shows the final orbital and mass distributions of our simulations, sub-divided by their initial disk profile.  The total integration time was 700 Myr, roughly corresponding to the start of the late heavy bombardment (Hartmann et al., 2000; Strom et al., 2005; Chapman et al., 2007; Bottke et al., 2012). We note that our simulations at 400 Myr -- the earliest likely start of the late heavy bombardment (Morbidelli et al., 2012) -- are qualitatively the same as at 700 Myr.

\subsubsection{The mass and orbital configuration of the planets}

It is clear from Figure 5 that steeper disks (higher $x$) produce smaller planets around 1.5 AU.  This is to be expected since steeper disks  have less mass in their outer parts (for a fixed disk mass). Therefore, they are closer to  the idealized  initial conditions proposed by Hansen (2009), namely a disk truncated at 1 AU. The simulations with $x$ of 2.5, 3.5 and 4.5  do not reproduce the terrestrial planets because they produce planets at around 1.5 AU that are systematically too massive compared to Mars\footnote{In fact, in one of the simulations with {\it x~}=~4.5 we observe the formation of a planet around 1.5 AU with mass very similar to Mars (actually slightly smaller). However, in this simulation there is another larger planet around 1.2 AU five times larger than Mars.}. Reasonable Mars analogs only formed in  simulations in our steepest disk ({\it x~}=~5.5). This case also produces good Earth and Venus analogs, and 20\% of {\it x~}=~5.5 simulations also formed good Mercury analogs. These Mercury analogs are usually leftover planetary embryos that started around 1 AU and were gravitationally scattered inwards by growing embryos.

Figure 6 shows the orbital configuration of the surviving bodies in our simulations. The final orbital distribution of simulated planets roughly matches that of the real terrestrial planets. To perform a quantitative analysis we make use of two useful metrics: the normalized angular momentum deficit (AMD; Laskar 1997) and the radial mass concentration (RMC; Chambers 2001) statistics. The AMD of a planetary  system measures the fraction of the planetary system angular momentum missing due non-circular and non-planar orbits.  The AMD is defined as (Laskar 1997): 

\begin{equation}
{\rm AMD}= \frac{{\sum_{j=1}^N}\Big[ m_j \sqrt{a_j} \> \Big( 1 - \cos i_j \> \sqrt{1-{e_j}^2}\> \Big)\Big]} 
{{\sum_{j=1}^N}\> m_j \sqrt{a_j}},
\end{equation}
where $m_j$ and $a_j$ are the mass and semi-major axis of planet $j$ and $N$ is the number of final bodies. ${\rm e_j}$ and ${\rm i_j}$ are the orbital eccentricity and inclination of the planet ${\rm j}$.

The RMC of a system measures how the mass in distributed in one region of the system. The value of RMC varies with the semi-major axes of planets (Chambers 1998, 2001; Raymond et al., 2009).  It is defined as: 
\begin{equation} 
{\rm RMC} = {\rm Max}\left(\frac{{\sum_{j=1}^N}\> m_j}{{\sum_{j=1}^N}\> m_j\big[\log_{10}\left(a/a_j\right)\big]^2}\right). 
\end{equation}
\noindent
In these formulas we consider as planets those objects  larger than 0.03 Earth masses and orbiting between 0.3 and 2 AU.

 Table 1 shows the mean number of planets formed in each set of simulations, the mean AMD, RMC  and water mass fraction (WMF). This table also provides the range over which these mean values were calculated. Steeper disks produce a larger number of planets per system, in agreement with Raymond et al. (2005). Steeper disks also tend to produce systems that are less dynamically excited. This is a consequence of two effects.  First, the initial individual embryo/planetesimal mass ratio in the inner regions of the disk is much higher for steeper disks (this ratio is about 40 [130] at 1 AU for {\it x~}=~2.5 [{\it x~}=~5.5], see Fig. 2), providing stronger dynamical friction, which tends to reduce the planets' final AMD (Raymond et al., 2006; O'Brien et al., 2006). Second, shallower disks have more mass in the asteroid belt and a more prolonged last phase of accretion.  This chaotic late phase excites the eccentricities (e.g., Raymond et al., 2014) of the surviving planets while reducing their number.  

 The final AMD of the simulated planetary systems is also a powerful diagnostic on the adequacy of the chosen initial mass fraction in planetesimals and individual mass ratio between planetary embryos and planetesimals in our simulations (effects of dynamical friction). Simulations of terrestrial planet formation have considered a range in the initial mass fraction of planetesimals, from 0 up to 50\% (e.g. Chambers, 2001; Raymond et al., 2004; O'brien et al., 2006; Izidoro et al., 2014; Jacobson \& Morbidelli, 2014). Given the moderate to low values of AMD of the protoplanetary systems in our simulations, even for the shallower disks, they suggest that our chosen initial mass fraction of planetesimals equal to 30-40\% is suitable.

Perhaps the most interesting and surprising result shown in Table 1 is that even our simulations considering {\it x~}=~5.5  were not able to reproduce the large RMC of the solar system terrestrial planets. The reason for this is that, although very steep disks have by definition a large radial concentration of mass in the inner regions, they tend to produce  a large number of planets between 0.3 and 2 AU. In fact, most of our simulations produce at least one extra planet between Mars and the inner edge of the asteroid belt ($\sim$ 2 AU). Simulation with {\it x~}=~5.5,  for example, produce on average 4.73 planets per simulation (of course, this is only possible because we have also a very low AMD for these systems which means that planets can stay very close to each other). These ``extra'' objects beyond Mars contribute to the observed  low values of RMC in these systems. However, even if we neglect planets between 1.6 and 2 AU when calculating the RMC, the mean value of this quantity in simulations with {\it x~}=~5.5 is 73 and only one planetary system showed an RMC larger than the solar system value of 89.9.

 In fact, no clear trend in RMC is observed for different values of {\it x} in Table 1. Moreover, simulations  with {\it x~}=~2.5 present a very wide range of RMCs. This may be consequence of stronger dynamical instabilities among forming terrestrial planets which eventually reduce the number of forming planets, sometimes achieving a larger RMC. As we have pointed out before, disks with shallow density profiles are more prone to instabilities because of smaller planetary-embryo/planetesimals individual mass ratio and of the protracted giant impact phase due to the interaction  with planetary embryos originally located in the asteroid belt region. As a result, these disks leads to a planetary system with a too large AMD compared to the real solar system. In summary, we find that no power-law disk profile is able to reproduce simultaneously the AMD and RMC of the real terrestrial planet system. All these results suggest that the high RMC of the terrestrial planets and their low  AMD  may be a signature of accretion in a very narrow disk with a  sharp disk edge truncation necessarily not far from 1 AU (Hansen, 2009), rather than in a power-law disk, even if very steep. Indeed, in Hansen's  simulations the final systems reproduced the RMC and AMD of the solar system  quite well. The same is true in the Grand-Tack simulations (Jacobson \& Morbidelli, 2014).

\begin{table*}
    \renewcommand*\arraystretch{1.5}
\begin{center}
\caption{Statistics analysis of the results of our simulations and comparison with the Solar System.\tablenotemark{a}}
\begin{tabular}{cccccc}
\tableline\tableline
$x$   & Mean N   & Mean AMD  &  Mean RMC  & Mean WMF          \\
\tableline
2.5        & 2.93 (2-4)     & 0.0047 (0.0003 - 0.0246)  & 60.25 (33.33 - 118.19)   & 2.64$\cdot10^{-3}$ (9.17$\cdot10^{-4}$-4.6$\cdot10^{-3}$)  \\
3.5        & 3.53 (2-5)     & 0.0020 (0.0005 - 0.0058)  & 50.40 (41.63 -  62.33)   & 9.60$\cdot10^{-4}$ (6.46$\cdot10^{-4}$-1.84$\cdot10^{-3}$)   \\
4.5        & 4.33 (2-6)     & 0.0012 (0.0001 - 0.0084)  & 52.23 (42.20 -  60.47)   & 3.11$\cdot10^{-4}$ (6.45$\cdot10^{-5}$-1.05$\cdot10^{-3}$)   \\
5.5        & 4.73 (4-6)     & 0.0004 (0.0001 - 0.0016)  & 57.15 (45.87 -  65.13)   & 2.81$\cdot10^{-5}$ (2.86$\cdot10^{-6}$-9.52$\cdot10^{-5}$)            \\   
SS         &  4                   & 0.0018                              & 89.9     &  $\sim10^{-3}$ [Earth\tablenotemark{b}] \\
\tableline
\end{tabular}
\footnotetext[1]{From left to right the columns are, the slope of the surface density profile, the mean number of planets, the mean AMD, mean RMC and mean WMF, respectively.}
\footnotetext[2]{The amount of water inside the Earth is not very well known. Estimates point to values ranging from 1 to $\sim$10 Earths' ocean (1 Earth's ocean is 1.4$\times10^{24}g$; see Lecuyer et al., 1998 and Marty 2012). A water mass fraction equal ${\rm 10^{-3}}$ assumes that 3 oceans are in Earth's mantle.}
\end{center}
\end{table*}

\subsubsection{Radial mixing and water delivery to Earth}

 We track the delivery of water from the primordial asteroid belt in to the terrestrial zone (inside 2 AU) by inward scattering of planetary embryos and planetesimals. For the initial radial water distribution in the protoplanetary disk we use the model widely used in classical simulations of terrestrial planet formation: planetesimals and protoplanetary embryos inside 2 AU are initially dry, those between 2 and 2.5 AU carry 0.1\% of their total mass in water and, finally, those beyond 2.5 AU have initially a water mass fraction equal to 5\% (e.g., Raymond et al., 2004; Izidoro et al., 2014). In our simulations, we assume no water loss during impacts or
through hydrodynamic escape (see Marty \& Yokochi 2006 and Matsui \& Abe 1986). The final water content of a planet formed in our simulations is the sum of its initial water content (if any) and the water content of all bodies that hit this respective object during the system evolution. Thus, for our initial water distribution, the water mass fraction of the planets in our simulations probably represents  an upper limit (e.g. Genda \& Abe, 2005).

The mean water mass fraction of the planets formed in our simulations are shown in Table 1. As expected, steeper disks produce drier planets, in agreement with Raymond et al. (2005). In fact, because of the setup of radial mass distribution (the total amount of mass is always equal to $\sim$2.5${\rm M_{\oplus}}$) of our simulations, steeper disks contain less mass in the outer part of the disk and consequently less water to be delivered to the terrestrial region. To compare quantitatively the results of our simulations with the amount of water  in the Earth we assume that  the Earth carries 3 oceans of water in the mantle\footnote{The total amount of water on Earths is  debated (Drake \& Campins, 2006). Estimates suggest a total amount of water in the Earth ranging between 1 and $\sim$10 Earth oceans (Lecuyer et al., 1998; Marty 2012).} (e.g. Raymond et al., 2009). Under this conservative assumption, only the flattest disks -- with {\it x~}=~2.5 and 3.5 -- are able to produce, on average, Earth analogs with Earth-like water contents. 
 
It is important to note that time zero of our simulations represent a stage of the formation of terrestrial planets after the gas disk dissipation. The radial mixing and water delivery in our simulations occurs purely due to gravitational interaction among bodies during the evolution of the system. However, recall that the main hypothesis of our work in based on the assumption that the solid mass distribution in the protoplanetary disk is dynamically modified during the gas disk phase by the radial drift of small particles due to gas drag, which can create pileups and very steep density profiles. In these circumstances, some level of radial mixing of solid material and also water delivery could have occurred in a very early stage (during the gas disk phase) because of drift of small particles (pebbles or small planetesimals) coming from large distances and entering into the terrestrial region. However, if the snowline had been in the vicinity of 1 AU, so to make the disk water-rich near the Earth region, it is difficult to understand why the giant planets formed far out and little mass formed in the asteroid belt (usually the formation of giant planet cores is attributed to snowline effects). Moreover, it would be strange that asteroids now resident in the inner asteroid belt (S-type, E-type bodies) are predominantly water poor. Our assumption that planetesimals were water-rich only beyond 2.5 AU was indeed inspired by the current compositional gradient of asteroids throughout the belt (e.g. DeMeo \& Carry, 2014).

\subsubsection{The last giant impact on Earth}

 We analyse in this section whether there is any trend or relationship between the timing of the last giant impact on Earth-analogs and the slope of the surface density profile. 

Radiative chronometer studies suggest that the last giant collision on Earth happened  between 30 and 150 My after the formation of the first solids\footnote{The time zero usually is assumed as the time of calcium-aluminium-rich inclusion's (CAIs) condensation.} in the nebula  (Yin et al., 2002; Jacobsen, 2005; Touboul et al., 2007; Taylor et al., 2009; Allègre et al., 2008; Jacobson et al., 2014).

\begin{figure}
\centering
\includegraphics[scale=.55]{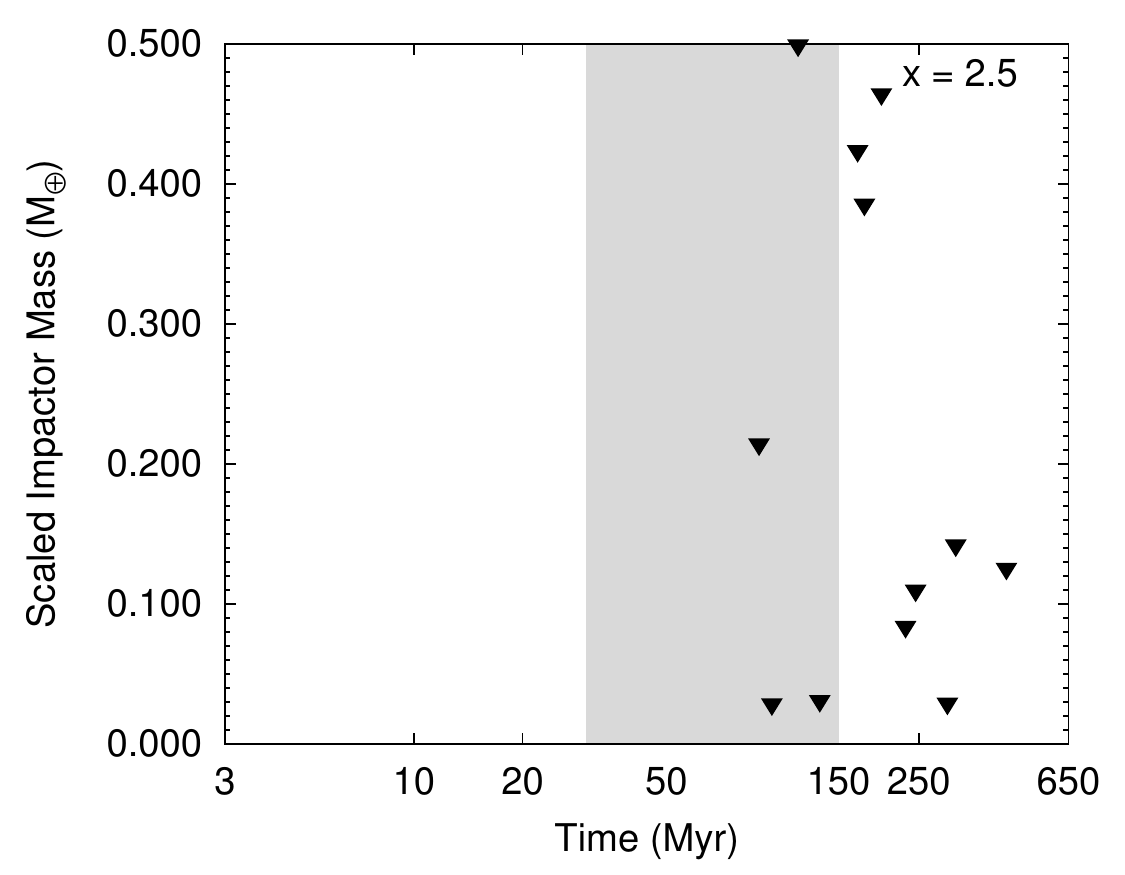}
\includegraphics[scale=.55]{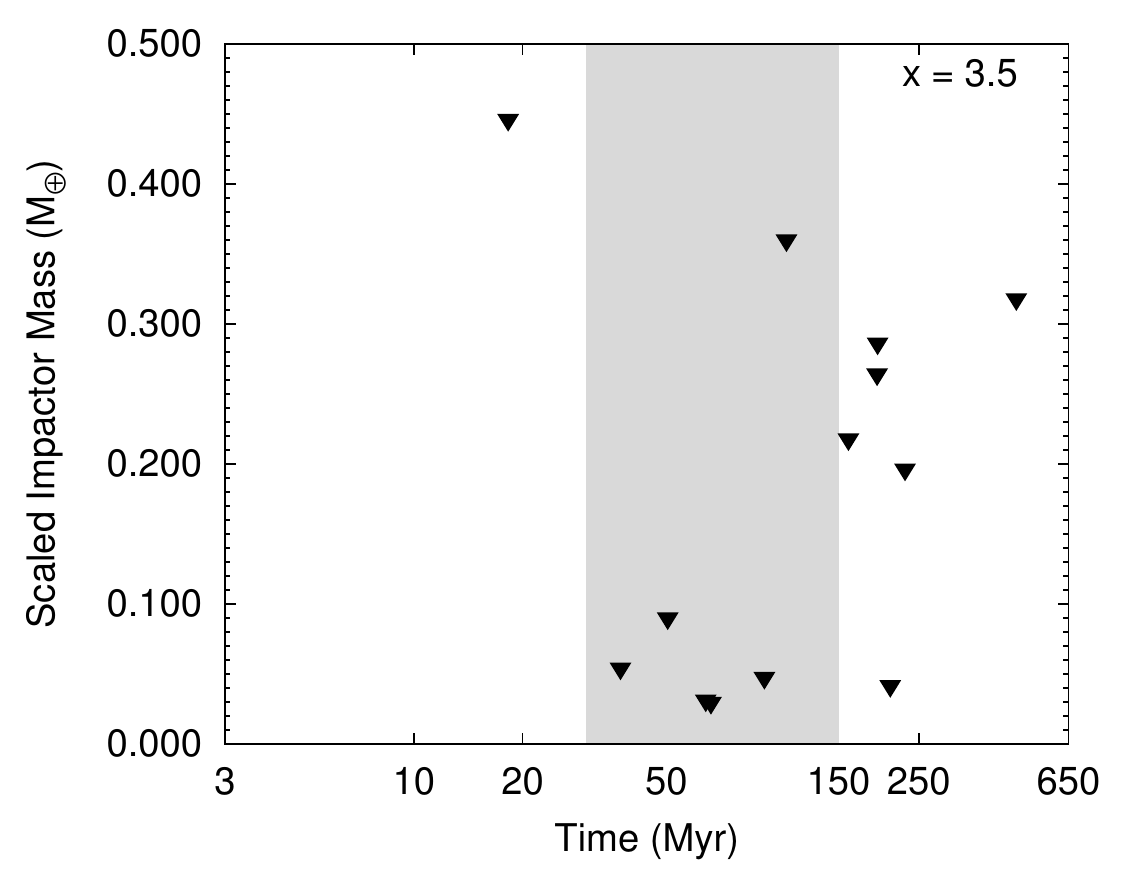}
\includegraphics[scale=.55]{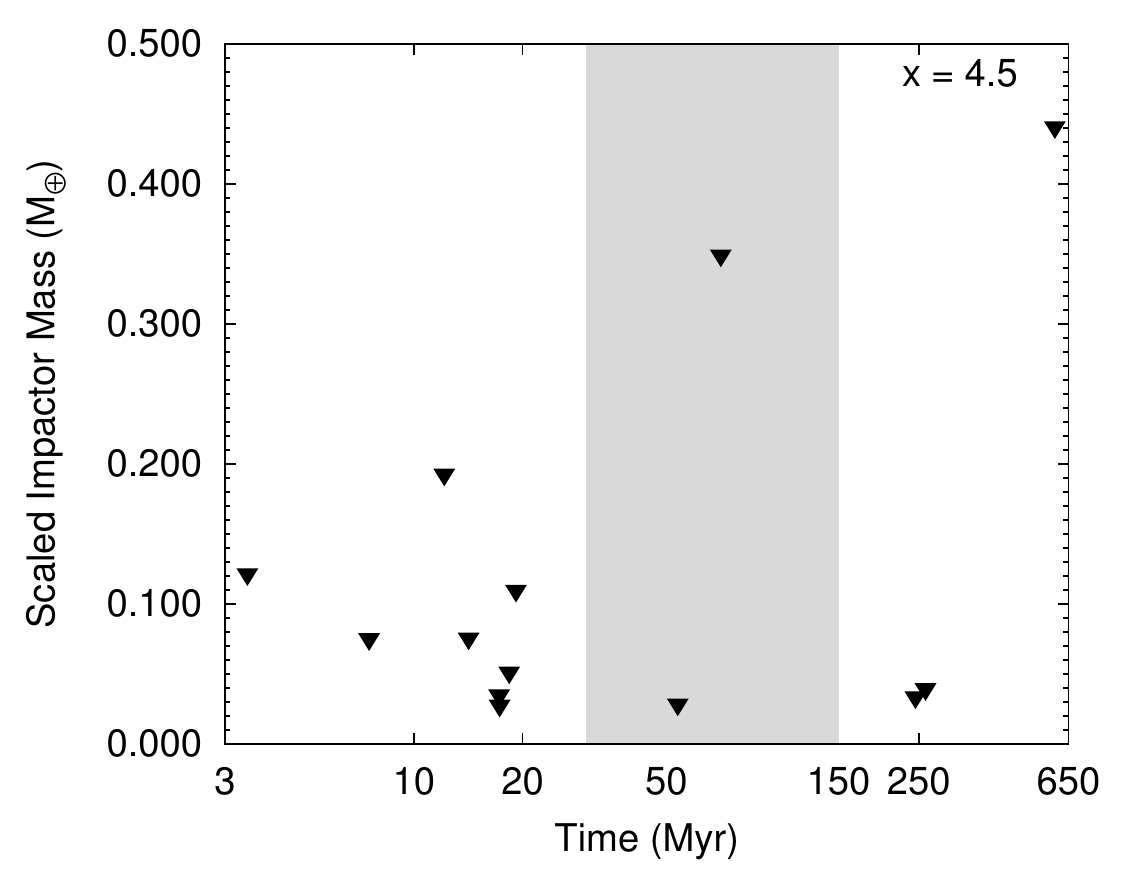}
\includegraphics[scale=.55]{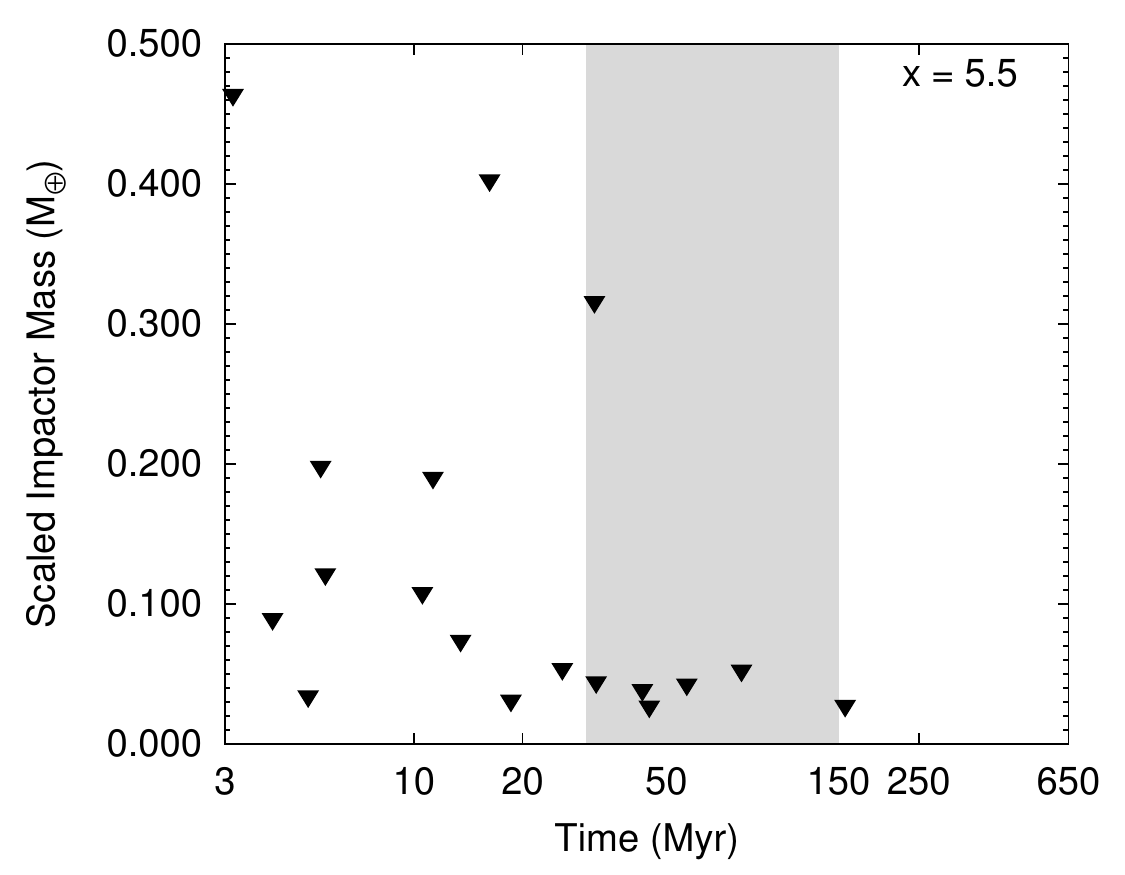}
\caption{Timing of last giant collision on Earth-analogs for all our simulations. The surface density profile is indicated on the right upper corner of each plot. The grey area shows the expected range (30-150 Myr) for the last giant impact on Earth derived from cosmochemical studies.}
\end{figure}

Obviously the time zero of our simulations  does not correspond to the time of the CAIs condensation. In our simulations, Jupiter and Saturn are assumed fully formed and the gaseous disk has already dissipated. Thus, it is likely that our scenario represents a stage corresponding  to $\sim$3 Myr after the time zero (e.g. Raymond et al., 2009). The analysis presented in this section already takes into account this offset in time.

To perform our analysis we define Earth-analogs as planets with masses larger than 0.6 Earth masses, orbiting between 0.7 and 1.2 AU and that suffer the last giant collision between 30 and 150 Myr (e.g. Raymond et al., 2009). We flag as giant collisions  those where the  scaled impactor mass is larger than 0.026 target masses (or $M_{\oplus}$). 

 Figure 7 shows the timing of last giant impact on all planets larger than 0.6 Earth masses and orbiting between 0.7 and 1.2 AU. There, we see a clear trend: planets forming in steeper disks are prone to have the last giant collision earlier. This is consistent with Raymond et al. (2005), who showed that disks with steeper surface density profiles tend to form planets faster. Applying our criterion to flag or not a planet as Earth-analog, we note that only 40\% of the Earth-mass planets formed in our simulations with {\it x~}=~5.5 (which can produce Mars analogs) have the giant impact later than 30 Myr. In many cases, the last giant collision happens much earlier than 30 Myr. Disks with {\it x~}=~2.5, on the other hand, show that about 75\% of the forming Earth-mass planets experience their last giant collision later than 150 Myr.

\subsubsection{Structure of the asteroid belt}

A successful model for terrestrial planet formation must also be consistent with the structure of the asteroid belt.  In fact, the asteroid belt provides a rich set of constraints for models of the evolution of the solar system (Morbidelli et al., 2015).  The present-day asteroid belt only contains roughly $10^{-3}$ Earth masses.  Figure 1 shows the current distribution of asteroids with magnitude smaller than $H<9.7$. This cutoff corresponds to objects larger than 50 km which safely make this sample free of observational biases (Jedicke et al., 2002). In addition, such large objects are unaffected for non-gravitational forces and they are unlikely to have been produced in family formation events.  As shown in Figure 1, the main belt basically spans from 2.1 to 3.3 AU and is roughly filled by orbital eccentricities from 0 to 0.3 and inclinations from 0 to 20 degrees.  Thus asteroids are, on average, far more excited than the terrestrial planets.  The challenge is to simultaneously explain the belt's low mass and orbital excitation. 

We now analyze the formation of the asteroid belt  in our N-body simulations.  Given the initial conditions that we have adopted for Jupiter and Saturn (different from the current orbits), our simulations implicitly assume that there will be a later instability in the giant planets' orbits. Thus, it is natural to wonder whether such a late instability would alter the final asteroid belt's structure.  The version of the Nice model that is consistent with observations is called the Jumping-Jupiter scenario (Morbidelli et al., 2009b; Brasser et al., 2009; Nesvorny \& Morbidelli, 2012), and invokes a scattering event between Jupiter and an ice giant planet (Uranus, Neptune or a rogue planet of comparable mass). In this model the orbital inclinations of the asteroids are only moderately altered (with a root-mean-square change of the inclination of only 4 degrees; Morbidelli et al., 2010). The subsequent 3.5 Gyr would have brought even smaller changes in the asteroid inclination distribution. Thus, in our analysis we compare directly the inclination and eccentricity distribution of the asteroids resulting from our simulations with the observed distribution, with the understanding that, if the simulated inclination distribution is too cold, it is highly unlikely that the subsequent evolution would reconcile the two distributions. To consolidate this claim, in Section 4.3, we perform numerical simulations to illustrate how the jumping-Jupiter scenario would affect the orbital distribution of our final systems. We note that the inverse is not necessarily true: a modestly over-excited asteroid belt could in principle be dynamically ``calmed'' during the Nice model instability by removal of the most excited bodies (Deienno et al., 2015).

Our main result is that only a subset of our simulations adequately excite the asteroid belt.  Figure 6 shows  the final orbital distribution of the protoplanetary bodies in all simulations after 700 Myr. For {\it x~}=~2.5, 3.5 and 4.5 planetary embryos as large as Mars were initially present in the asteroid belt (see Fig 2), and many of these survive in the main belt (see also Figure 5). These planetary embryos interact gravitationally with planetesimals increasing planetesimals' velocity dispersions and generating higher eccentricities or/and inclinations.  Of course, it is inconsistent with observations for large embryos to survive in the main belt because they would produce observable gaps in the resulting asteroid distribution (O'Brien et al., 2006; Raymond et al., 2009).  This indeed happens for {\it x~}=~2.5, 3.5 and 4.5.

For {\it x~}=~2.5, most of the bodies that survived between 2 and 3 AU are large planetary embryos. Planetesimals that survived in this region have very high orbital inclinations and orbital eccentricities above 0.1. Only the very outer part of the main belt contains a significant planetesimal population.  The situation is similar for simulations with {\it x~}=~3.5.  These cases provide a reasonable match to the orbital distribution of bodies contained within the belt but the existence of planetary embryos makes them inconsistent with today's asteroid belt.  

In the simulations with {\it x~}=~4.5 many planetary embryos survived interior to 2.5 AU.  The region exterior to 2.5 AU is mostly populated by planetesimals (and some $<$Moon-mass planetary embryos; see Figure 5) with a broad range of excitations, although there is a modest  deficit of high-inclination planetesimals in the outer main belt.  This set of simulations provides the closest match to observations.  Remember, however, that in addition to the remaining long-lived large planetary embryos frequently observed in the asteroid belt (Fig. 5 and 6), in this disk planets around 1.5 AU are systematically about 5 times more massive than Mars (Fig. 5).

The simulations with {\it x~}=~5.5 (the only ones that reproduce the small mass of Mars) suffer from a severe under-excitation of the asteroid belt (Fig 6).  Planetesimals initially between 2.1 AU and 2.5 AU gain eccentricities up to 0.2-0.3 and inclinations up to 15 degrees. However, planetesimals beyond 2.5 AU are characterized by $e < 0.1$ and $i < 10^\circ$.  Indeed, the typical inclination in the outer belt is less than $3^\circ$. This is certainly inconsistent with the real population of asteroids. We stress that any initial disk mass distribution which is not a power law but predicts less mass in the asteroid belt than our disk with surface density profile proportional to ${\rm {\it r}^{-5.5}}$ (e.g. Hansen, 2009) would obviously have the same problem.

 We now quantitatively compare the results of our simulations with the real asteroid belt. Table 2 shows the root-mean-square of the orbital eccentricity (${e_{{\rm rms}}}$) and inclination ($i_{{\rm rms}}$) of real and simulated populations.  

To perform our analysis we divided the real main belt in two sub-populations and calculated the quantities (${ e_{{\rm rms}}}$ and ${  i_{{\rm rms}}}$) in each. Bodies belonging to the subpopulation A have semi-major axis between 2.1 AU and 2.7 AU and those form the subpopulation B have semi-major axis ranging from 2.7 AU to 3.2 AU. We divide the simulated belt populations following these same criteria. We only included simulated planetesimals in the asteroid belt region for which the orbital eccentricity is smaller than 0.35, the orbital inclination is smaller than 28 degrees (see Figure 1) and which are confined within either region A or B. In principle, these  cutoff values were chosen using as inspiration the ``edges'' of the  real population of asteroids in the main belt with magnitude H$<$9.7 (see Fig 1). Planetesimals  with {\it e} or {\it i} (significantly) higher than those of the real population are eventually removed from the system due to close encounters with the giant planets or with the forming terrestrial planets. This is observed in our simulations. For example, we computed the fraction  of planetesimals between 2.1 and 2.7 AU  that have orbital eccentricity and inclination higher than our cutoff values, at two different snapshots in time, at 200 Myr and 700 Myr. Our results show that, at 200 Myr, 40\%  and 27\% of planetesimals between 2.1 and 2.7 AU have eccentricities larger than 0.35 in simulations with {\it x~}=~2.5 and 3.5, respectively. At 700 Myr, however, for disks with {\it x~}=2.5 this number is reduced to 0\% (meaning that there is no planetesimal with orbital eccentricity higher than our cutoff value -- they have been mostly ejected from the system or collided with the star). For {\it x~}=~3.5 this number drops to 20\% at 700 Myr. Thus, ignoring such bodies in our analysis is adequate since we avoid biasing our conclusions in this sense.

 Table 2 also shows the size of our samples, i.e., the sizes of the populations of real and simulated asteroids composing the regions A and B  in our analysis. For robustness, our analysis of the (${ e_{{\rm rms}}}$) and (${ i_{{\rm rms}}}$) is presented together with a statistical test. We use the Anderson-Darling (AD) statistical test, a more sensitive and powerful refinement of the popular Kolmogorov-Smirnov (K-S) statistical test (Feigelson \& Babu, 2012). The AD test is often used to decide whether two random samples have the same statistical distribution. A low probability (or p-value) means that the samples are dissimilar and unit probability means they are the same. We apply the AD test separately to subpopulations A  and  B and use a significance level to discard the similarity among samples smaller than 3\%.

Table 2 summarizes our analysis.  Before comparing the simulated and real belt populations let us compare regions A and B of the real belt population. The ${ e_{{\rm rms}}}$ of asteroids in regions A and B of the real main belt are similar. The same trend is observed in ${ i_{{\rm rms}}}$ values. The excitation level over the entire asteroid belt is equally balanced, providing a strong constraint for our simulations. Because the final number of objects  in regions A and B are relatively small for those simulations with {\it x~}=~2.5 and 3.5 (after 700 Myr of integration) we also apply our statistical tests using the system dynamical states at 300 Myr. Around 300 Myr, the number of planetesimals in the asteroid belt region is higher than that at 700 Myr but most of the planets are  almost or fully formed (e.g. Figure 3) which supports this choice. This secondary analysis is also certainly important in the sense of certifying that our statistical analysis are not biased by the smaller number of surviving objects in these systems.

 The level of eccentricity excitation of the planetesimals in the A region, at both 300 Myr and 700 Myr, are comparable for simulations with {\it x~}=~2.5, 3.5 and 4.5. This result is counter-intuitive since it does not reflect the fact that shallow disks have more mass and larger planetary embryos in the asteroid belt and so should dynamically excite planetesimals more strongly than steeper disks (Figure 5 and 6).  We do indeed find that simulations with shallower disk profiles have a larger proportion of planetesimals that are over-excited with respect to our asteroid belt $e$ and $i$ cutoffs, and a faster decay in this population.  However, the rms excitation of the planetesimals within the defined belt  remains (for some disks; see Table 2) similar.  We suspect that this may be linked with a combination of small number statistics and dynamical excitation by remnant planetary embryos in the belt.

 The AD test applied to the A region of our simulations with {\it x~}=~2.5, 3.5 and 4.5  does not reject the hypothesis that these populations have the same statistical distribution than the real belt. In fact the p-value in these cases are about 68\%, 7\% and 22\% for  simulations with {\it x~}=~2.5, 3.5 and 4.5, respectively. However, the disk with {\it x~}=~5.5 is dynamically much colder than the real population. The AD test rejects with great confidence the possibility that this simulated population of asteroids match the real one.

 As may be expected, there are some modest differences for the values of ${ e_{{\rm rms}}}$ and ${i_{{\rm rms}}}$ (and consequently the p-values)  shown in Table 2 at 300 Myr and 700 Myr. It is natural to expect that a system may be dynamically excited over time as far there is a mechanism to do so. However, we also  note that a system may be dynamically ``calmed'' if a fraction of the objects with (very) eccentric/inclined orbits are removed from the system while those with dynamically colder orbits tend to be kept in the system. This is observed, for example, for our simulations with {\it x~}=~2.5 (the ${ e_{{\rm rms}}}$ of the A and B regions are smaller at 700 Myr than at 300 Myr). However, to summarize, none of our disks had its dynamical state altered from 300 Myr to 700 Myr up to the point that the simulated asteroid belt match the real one.

Our simulated population of outer main belt (region B) asteroids in disks with {\it x~}=~2.5, 3.5 and 4.5 have eccentricity distributions that match the real outer belt. However, the B population of our disks with {\it x~}=~5.5, are statistically under-excited compared with the real belt.   

The real asteroid belt's inclination distribution turns out to be the most difficult constraint for the simulations, in particular the similar ${i_{{\rm rms}}}$ values of the inner and outer main belt.  Our simulations with {\it x~}=~2.5, 3.5 and 4.5 produce inner main belts (region  A) with inclinations that are statistically over-excited compared with the real one.  The simulation with {\it x~}=~5.5, on the other hand, drastically under-excites the inclinations of the inner main belt.  In the outer main belt (region B), only the simulations with {\it x~}=~3.5 provide a reasonable match to the actual asteroid inclinations.  Simulations with steeper surface density profiles ({\it x~}=~4.5 and 5.5) under-excited the outer main belt's inclinations while simulations with shallower profiles ({\it x~}=~2.5) over-excited the inclinations.  

To summarize, none of our simulations produced an asteroid belt with the same level of dynamical excitation as the present-day main belt.  Of course, we need to keep in mind that the end-state of our simulations does not correspond to the present-day.  Rather, it only corresponds to the start of the late instability in the giant planets' orbits envisioned by the Nice model, which certainly acted to modestly deplete and reshuffle the asteroid belts' orbits (Morbidelli et al., 2010; Deienno et al., 2015).  In addition, there are other potential mechanisms of excitation that we have not yet considered.

\begin{table*}
\caption{ Statistical analysis of the orbital architecture of the asteroid belt. Comparison of the root mean square of the orbital eccentricities and inclinations of the real and simulated populations of asteroids and use of the Anderson-Darling statistical test to compare the different populations.}              
\label{table:1}      
\centering                                      
\begin{tabular}{c|| c| c| c| c| c| c| c}     
\hline                        
 \multicolumn{2}{c|}{}   & \multicolumn{2}{|c|}{${\it e}_{{\rm rms}}$} & \multicolumn{2}{|c}{$i_{{\rm rms}}$ [deg]}  & \multicolumn{2}{|c}{{\it N}}\\   

\hline                                   
          & Region &  2.1-2.7 AU [A]   &  2.7-3.2 AU [B] &  2.1-2.7 AU [A]     &  2.7-3.2  AU [B]  &  2.1-2.7 AU [A]     &  2.7-3.2  AU [B] \\
          \hline\noalign{\vskip .03in}        
          \hline       
\multirow{2}{*}{Real population} &\multirow{2}{*}{ } &\multirow{2}{*}{ 0.18}         &  \multirow{2}{*}{0.16  }          &   \multirow{2}{*}{11.2} & \multirow{2}{*}{12.6 } &   \multirow{2}{*}{173} & \multirow{2}{*}{352}      \\
  & &  & & & & & \\
\hline\noalign{\vskip .03in} 
\hline
  power index [{\it x}]/Time     &   \multicolumn{6}{|c}{}  \\
\hline
\textcolor{gray}{2.5/300 Myr}  & & \textcolor{gray}{0.21 (0.07)}            & \textcolor{gray}{0.17 (0.23)} & \textcolor{gray}{20.1 (0)}   & \textcolor{gray}{16.4 (${\rm 10^{-4}}$)} &  \textcolor{gray}{33} & \textcolor{gray}{59} \\    
2.5/700 Myr  & & 0.19 (0.68)            & 0.13 (0.18) &  19.3 (${\rm 10^{-5}}$)  & 16.5 (0.01) & 13 & 18 \\    
\textcolor{gray}{3.5/300 Myr}  & & \textcolor{gray}{0.21 (0.01 )}           & \textcolor{gray}{0.19 (${\rm 10^{-3}}$)} & \textcolor{gray}{18.3 (0)} &  \textcolor{gray}{12.1 (0.67)} & \textcolor{gray}{64} & \textcolor{gray}{84}  \\ 
3.5/700 Myr  & & 0.22 (0.07) & 0.15 (0.17) & 19.7 (${\rm 10^{-5}}$) &  12.8 (0.29) & 10 & 28 \\      
\textcolor{gray}{4.5/300 Myr}  & & \textcolor{gray}{0.20 (0.17)}           & \textcolor{gray}{0.15 (${\rm 10^{-3}}$)} & \textcolor{gray}{12.3 (0.07)} &  \textcolor{gray}{7.13 (0.0)} & \textcolor{gray}{189} & \textcolor{gray}{277}  \\ 
4.5/700 Myr  & & 0.17 (0.22)            & 0.15 (0.13) &  13.4  (${\rm 10^{-3}}$)  &  7.8 (0.0)  & 55  & 162 \\   
\textcolor{gray}{5.5/300 Myr}  & & \textcolor{gray}{0.12 (0.0)}           & \textcolor{gray}{0.04 (0.0)} & \textcolor{gray}{6.9 (0.0)} &  \textcolor{gray}{2.2 (0.0)} & \textcolor{gray}{546} & \textcolor{gray}{623}  \\ 
5.5/700 Myr  & & 0.13 (0.0) & 0.05 (0.0) & 7.8  (0.0)   &  2.2 (0.0) & 382 & 620 \\   
\hline
\multicolumn{8}{l}{%
  \begin{minipage}{17cm}%
  \vspace{0.2cm}
\scriptsize {The columns are:  root-mean-square eccentricity (${{\it e}_{{\rm rms}}}$), inclination (${{\it i}_{{\rm rms}}}$) and number ({\it N}) of planetesimals (or asteroids). Each of these columns are divided in two sub-columns showing the range  of semi-major axis of the asteroids  taken in each sample for the calculation of  ${ e_{\rm rms}}$, ${i_{\rm rms}}$ and N. The respective subpopulations are called A which contains only those bodies orbiting with semi-major axis between 2.1 AU and 2.7 AU and subpopulation B which contains bodies orbiting with semi-major axis between 2.7 AU and 3.2 AU. In the left, we have the real and simulated population of asteroids (power index [{\it x}]). In the first column from left, we also show the corresponding time, representing a snapshot of the system dynamical state, which our statistics are computed.  Each table entry in parentheses is the p-value of the respective simulated sample when compared with the real population (Figure 1) using the Anderson-Darling Statistical test.}
  \end{minipage}%
}\\                                          
\end{tabular}

\end{table*}

\section{Mechanisms of dynamical excitation of the asteroid belt}

As shown before, the only disk capable of producing a Mars-mass planet around 1.5 AU results in an asteroid belt with an orbital distribution way too cold compared to the real one. Complementing the analysis disclosed previously,   we present in Section 4.1 the results of numerical simulations including N-body self interacting planetesimals and results of semi-analytical calculations describing the gravitational stirring among a large number of planetesimals.

\subsection{Self-interacting planetesimals in the asteroid belt}

The key open question at this point in our study is whether, in the absence of planetary embryos, the asteroid belt can self-excite.  This is the scenario represented by our simulations with {\it x~}=~5.5.  In those simulations the region beyond 2.5 AU contains many planetesimals but they have very low eccentricities and inclinations.  However, the simulations did not include interactions between planetesimals, which should act a source of viscous stirring to increase the orbital excitation. We tackled this problem using both an N-body and a particle-in-a -box approach.

\subsubsection{N-body numerical simulations}

In this scenario, we assume that the region of the disk beyond 2 AU is populated only by planetesimals. We performed a simulation containing 300 asteroids with individual masses of $1.3\times10^{-4} M_\oplus$. This translates to objects with diameters of about $\sim 1000$ km.  The total mass is $0.04 M_\oplus$, which is about 80 times the current belt mass.  We uniformly distributed these objects between 2 and 3.5 AU with initially circular orbits and inclinations chosen randomly from the range of $10^{-4}$ to $10^{-3}$ degrees (mean anomalies and longitudes of ascending nodes were randomized). Jupiter and Saturn have the same orbital configuration as our previous simulations. We have numerically integrated the dynamical evolution of these bodies for 200 Myr.

\begin{figure}
\centering
\includegraphics[scale=.7]{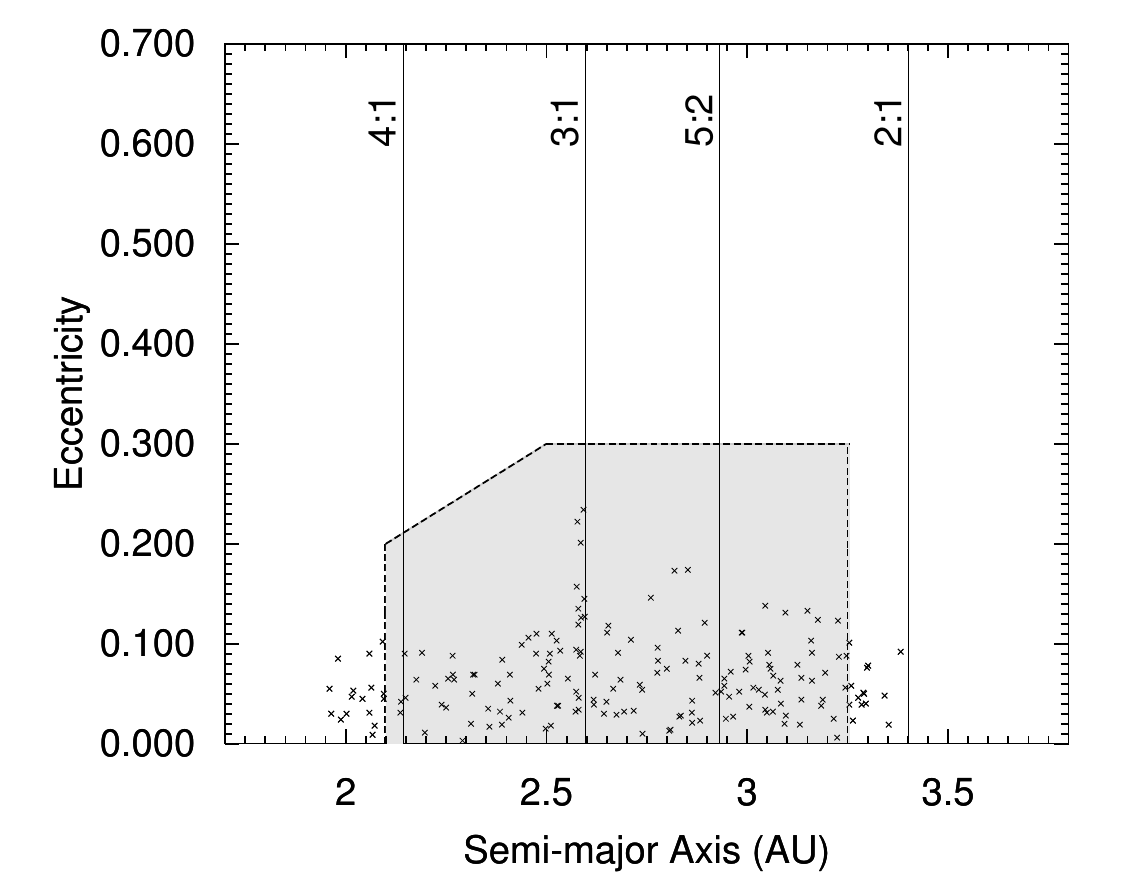}
\includegraphics[scale=.7]{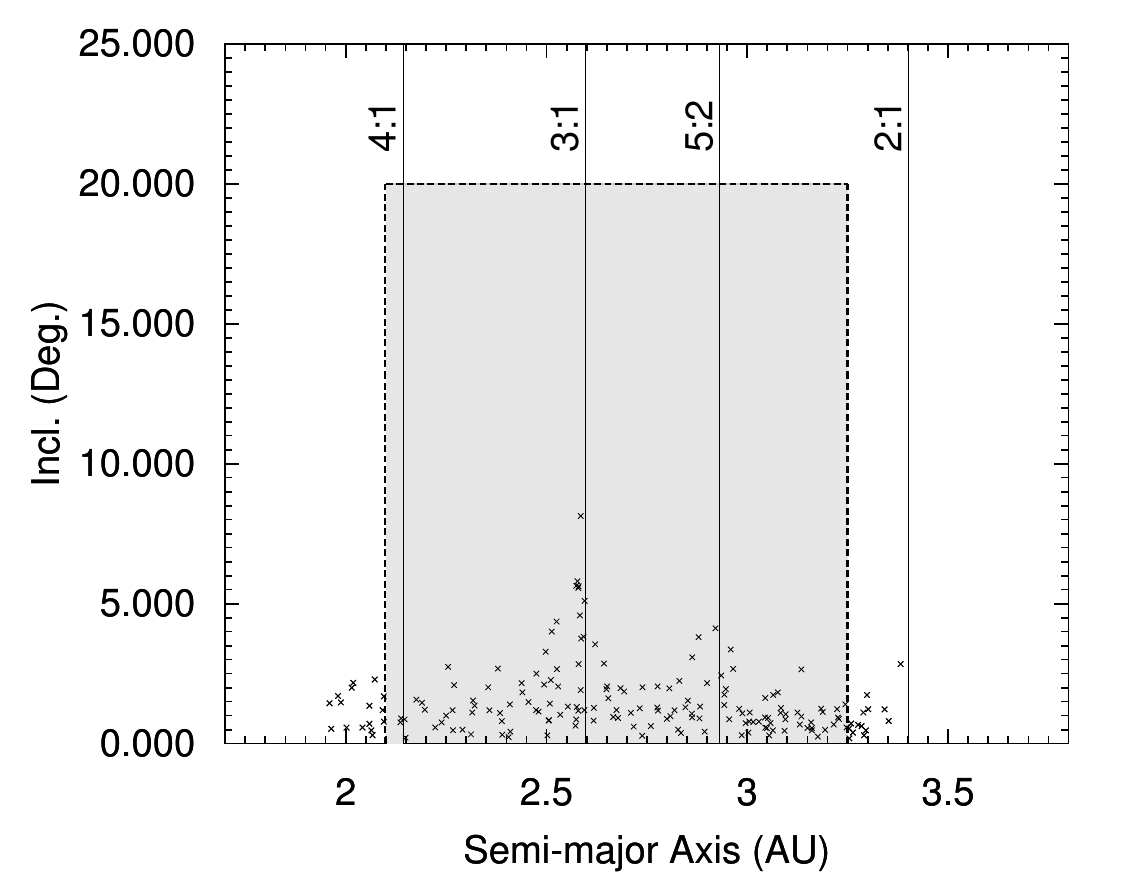}
\caption{Final state of one simulation considering self interacting planetesimals distributed between 2 and 3.5 AU after 200 Myr of integration. Each planetesimal have an initial small mass equal to $1.3\times10^{-4} M_\oplus$. The upper plot shows eccentricity versus semi-major axis. The lower plot shows orbital inclination versus semi-major axis. The vertical lines show the locations of mean motion resonances with Jupiter.}
\end{figure}

Figure 8 shows the self-excitation of the simulated asteroid belt after 200 Myr.  Orbital eccentricities of asteroids are excited up to about 0.1 and orbital inclination to no more than 2-4$^\circ$. And these values  of eccentricity and inclination should be considered as upper limits, because the self-excitation does not lead to a substantial loss of bodies, as it is the case here, and the belt could not contain so much mass in large asteroids. Thus, the real asteroid self-excitation would have been even smaller.

\subsubsection{Semi-analytical calculations: Particles in a box approximation}

We also performed an analysis using the particle in a box approximation  (e.g. Safronov, 1969; Greenberg et al. 1978; Wetherill and Stewart 1989, 1993; Weidenschilling et al., 1997; Kenyon and Luu 1998;  Stewart \& Ida, 2000; Morbidelli et al., 2009a).  We used a very similar setup to the N-body simulations.  Our calculation describes the gravitational stirring among a large number of planetesimals in the region of the asteroid belt, using the Boulder code (Morbidelli et al., 2009a). As in the previous case the individual masses of planetesimals are purposely chosen to represent an extreme case compared to  the expected size frequency distribution for the primordial population of asteroids in the region (Bootke et al., 2005; Morbidelli et al., 2009a; Weidenschilling, 2011). Our dynamical system is composed by a one-component population of 1000 Ceres-mass planetesimals with diameter of about 1000 km distributed from 2 AU to 3.2 AU. The orbital eccentricities and inclinations of these bodies are set initially to be equal to $e_0=0.0001$ and  $i_0\simeq2.8\cdot 10^{-3}$ degrees. We follow the evolution of the root mean square eccentricity and inclination of this population of objects  for 4 Gyr, nearly the age of the solar system. 

Figure 9 shows the outcome of this calculation.  As expected, the root mean square eccentricity and inclination of this population is much smaller than that of the real population of the bodies in the asteroid belt. The $e_{rms}$ of this populations stays below 0.1 while the orbital inclination is smaller than $3^\circ$. This is a robust result given that the amount of mass carried out by 1000 Ceres mass is about $\sim$0.2 Earth masses. However,  the formation of Mars  analogs around 1.5 AU is only possible in those disks as steep as ${\rm {\it r}^{-5.5}}$. But the amount of mass  between 2 AU and 3.2 AU in these disks (see Section 2) is only about $\sim$0.03 Earth masses. Thus, the results from Figure 9 are firm upper limits on the disk's self-excitation.

It is important also to stress that  in our semi-analytical calculation we neglect mass accretion, fragmentation and erosion due to collisions between planetesimals. If taken into account, these effects in general should only slightly change the evolution of ${\rm e_{rms}}$ and  ${\rm i_{rms}}$  of this population (e.g. Stewart \& Ida, 2000). However, their effects could become important in cases where there is mass growth and formation of very large bodies during the  timescale of the evolution of the system. In this case, our calculation would probably underestimate $e_{rms}$ and $i_{rms}$ since we are using an one-component population of planetesimals (during all the 4 Gyr the planetesimals have the same mass). In our application, however, this seems to be an adequate approach since we do not see  in the asteroid belt today any body larger than Ceres.

Of course, our statistical calculation does not take into account the gravitational perturbation from the giant planets. However, because they have initially almost coplanar and circular orbits their secular effects on the planetesimals is very small (see Raymond et al., 2009 and our Figure 8). In this case, only those bodies near mean motion resonances should gain modest inclinations and eccentricities.

\begin{figure}
\centering
\includegraphics[scale=.8]{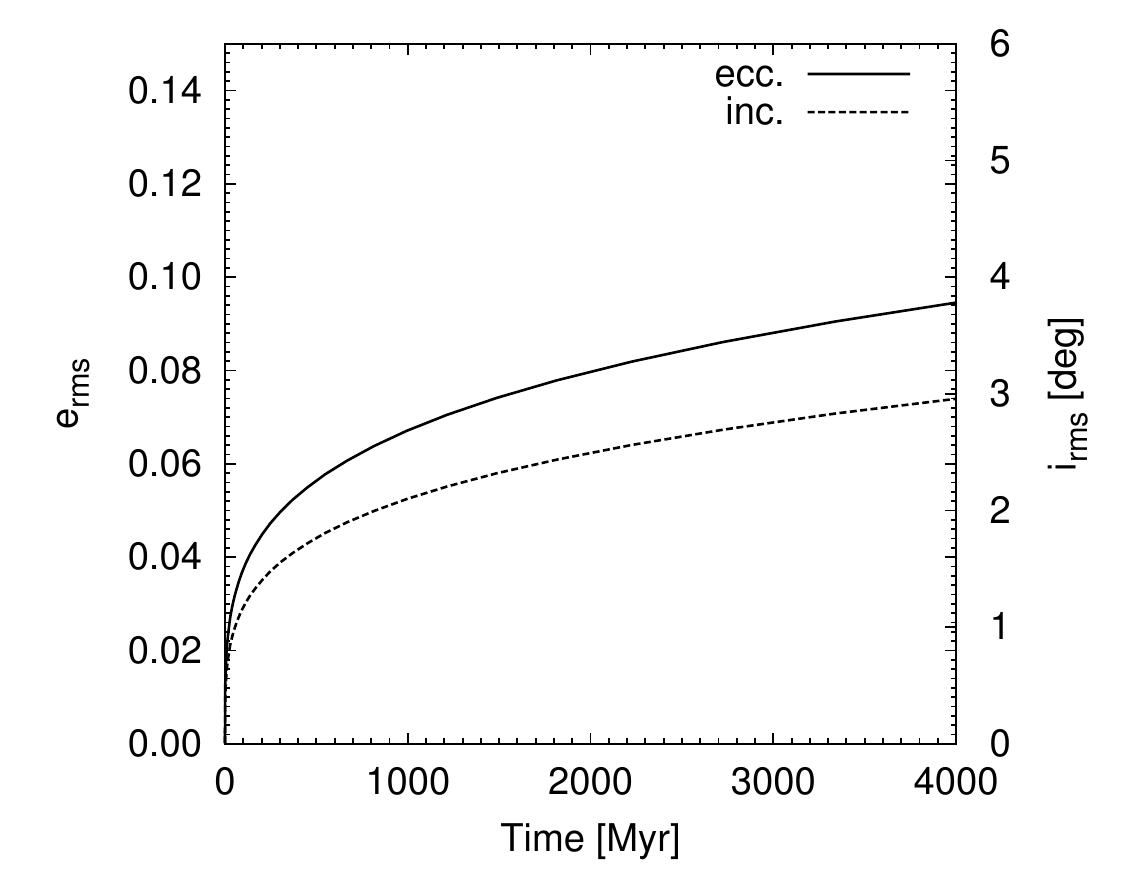}
\caption{Evolution of root mean square eccentricity and inclination of a population of 1000 Ceres mass planetesimals for 4 Gyr obtained from statistical calculations.}
\end{figure}

\subsection{Late dynamical instabilities and migration of giant planets}

 The giant planet configuration assumed in our simulations represents an epoch prior to the late stage of dynamical instabilities of the giant planets. In our simulations, only the very steepest power-law disk (with {\it x~}=~5.5) formed viable Mars analogs, but the asteroid belts in these simulations were under-excited compared with the present-day belt.  But could the later stage of dynamical instability among the giant planets solve this issue? To address this question we performed simulations to analyze how the final state of the planetary systems produced in disks with {\it x~}=~5.5 is affected after the migration of Jupiter and Saturn within the ``jumping-Jupiter'' scenario (Morbidelli et al., 2009b; Brasser et al., 2009; Nesvorny \& Morbidelli, 2012). Specially, we focus on the evolution of the orbital inclination of the asteroid belt.

To mimic the evolution of Jupiter and Saturn in the jumping-Jupiter scenario, for simplicity, we assume that Jupiter and Saturn instantaneously evolve from the final state of our simulations of terrestrial planet formation, where they still have  almost coplanar and circular resonant orbits (at {\it t~}=~700 Myr) to their current orbits. We use this approach for two reasons. First, it is computationally cheap. Second, this assumption avoids many difficulties that would emanate from using a  forced migration for these planets. A poorly-controlled forced migration for Jupiter and Saturn could bias our analysis and conclusions. In contrast, this approach allows us to assess the essential signatures printed in the asteroid belt by the jumping-Jupiter. Simulations of the jumping-Jupiter process (Nesvorny \& Morbidelli, 2012) show that it is unlikely that Jupiter passed through a dynamical phase with an inclination significantly larger than its current one. Thus, substituting the original jovian orbit with quasi-null inclination with the current orbit, as we do here, is a good approximation for the jumping-Jupiter evolution, at least as far as the inclination excitation is concerned.

\begin{figure}
\centering
\includegraphics[scale=.8]{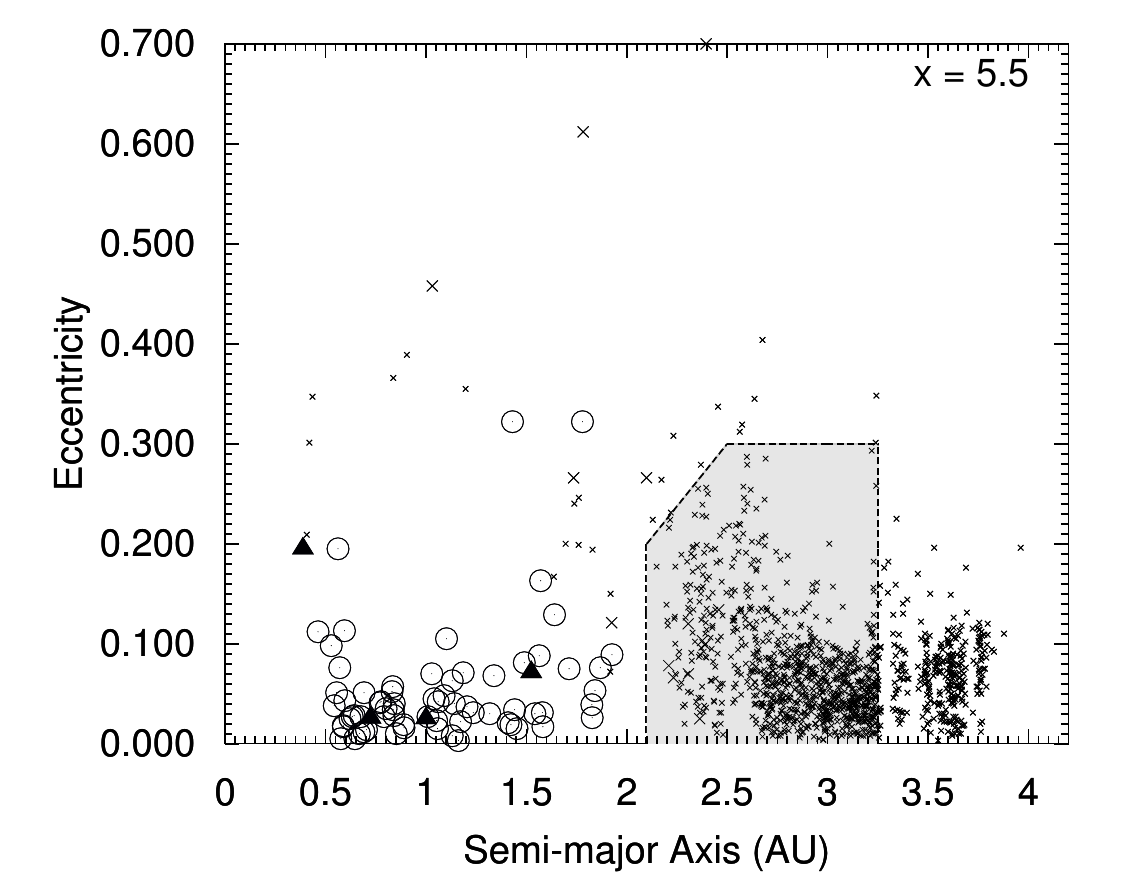}
\includegraphics[scale=.8]{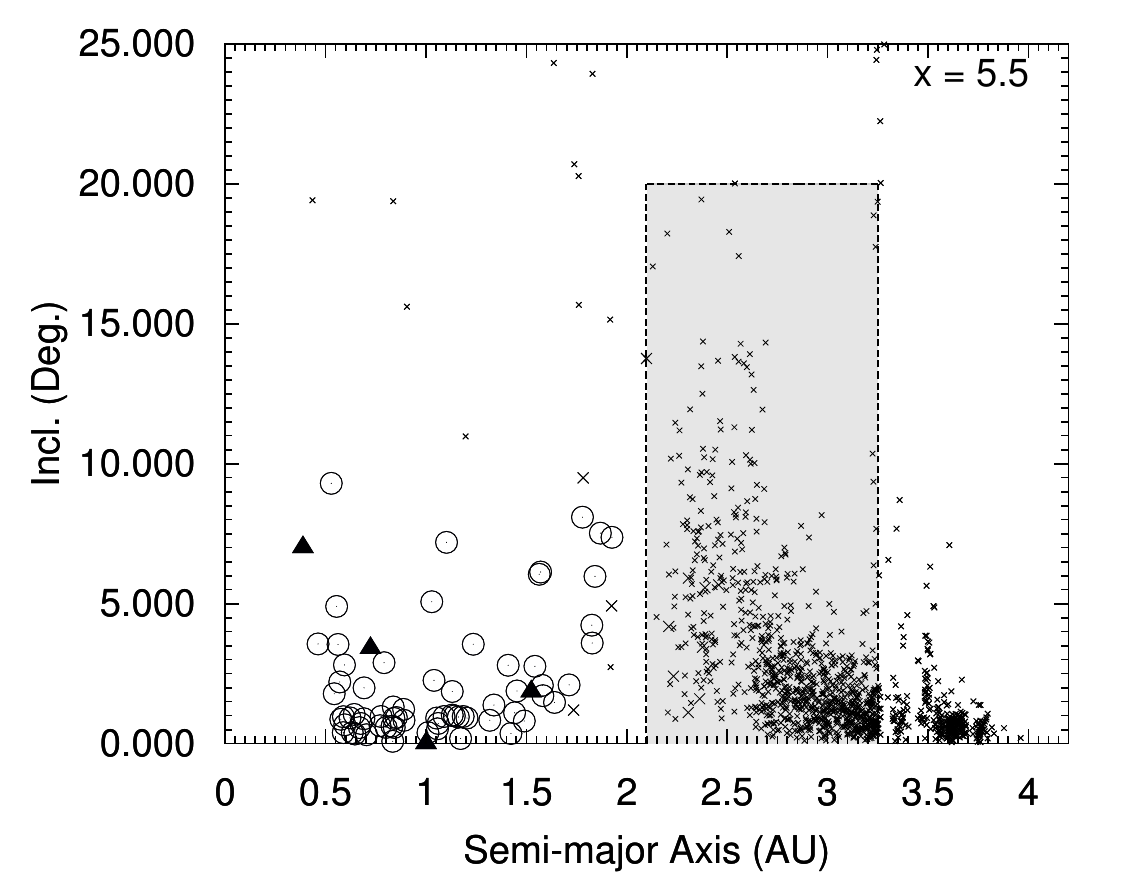}
\caption{Same that Figure 6 but after the jumping-Jupiter like evolution of Jupiter and Saturn.}
\end{figure}

Figure 10 shows the final state of all our simulations, with {\it x~}=~5.5, 30 Myr after the jumping-Jupiter like evolution of Jupiter and Saturn. We note that the final orbital distribution of bodies after the jumping-Jupiter like evolution of Jupiter and Saturn does not change qualitatively. In fact, the strongest effect appears in the orbital eccentricity distribution, but still planetesimals beyond 2.5-2.8 AU have eccentricities $\sim$0.1 or smaller. The orbital inclinations of bodies beyond 2.5-2.8 AU also are only weakly affected.

We stress that even if Jupiter and Saturn had reached their current orbits in a much smoother and slow migration fashion (Tsiganis et al., 2005; Minton \& Malhotra, 2009; 2011) than predicted by the Jumping-Jupiter scenario, the orbital inclination distribution of simulated bodies beyond $\sim$2.8 AU (simulations with {\it x~}=~5.5) would remain very dynamically cold. This is because planetesimals beyond 2.8 AU would preserve their initial inclination distribution because they are not greatly affected by the sweeping of secular resonances  during the giant planet migration (e.g. Figure 4 in Morbidelli et al., 2010). In contrast, most of the planetesimals/asteroids inside 2.8 AU would have their orbital inclinations increased significantly. Slow migration of giant planets also comes with the drawback of implying a slow sweeping of secular resonances across the asteroid belt and terrestrial region which likely make the terrestrial planets too dynamically excited in the end (Brasser et al., 2009;  Agnor and Lin, 2012).

\subsection{Effects of the initial orbits of the giant planets.}

 In this paper we have assumed that Jupiter and Saturn were on nearly circular and coplanar orbits during the late stages of terrestrial accretion.  Yet the giant planets' orbits at early times are poorly constrained.  There are two qualitatively different schools of thought (see recent reviews by Morbidelli et al., 2012, Raymond et al., 2014, and Raymond \& Morbidelli 2014 for more detailed discussions).

The first model proposes that, when the gaseous protoplanetary disk dissipated, the giant planets were stranded on orbits sculpted by planet-disk interactions.  The most likely orbital configuration is a low-eccentricity, low-inclination 3:2 mean motion resonance between Jupiter and Saturn (Masset \& Snellgrove 2001; Morbidelli \& Crida 2007; Pierens \& Nelson 2008; although a 2:1 resonance is also possible -- Pierens et al., 2014).  The giant planets acquired their current orbits during a late phase of orbital instability, i.e. the Nice model (Morbidelli et al., 2007; Levison et al., 2011; Batygin et al., 2010; Nesvorny and Morbidelli, 2012).  Within a standard, $\Sigma \propto {\it r}^{-1.5}$ disk, this model systematically produces unrealistically large Mars analogs (Chambers 2001; Raymond et al., 2009; Morishima et al., 2010). This problem motivated the development of the Grand tack scenario (Walsh et al., 2011).

The second model proposes that Jupiter and Saturn were on or close to their current orbits from very early times (we refer the reader to Raymond et al., 2009 for a discussion), prior to the terrestrial planet accretion phase (e.g. Chambers, 2001). This could be the case if the giant planet instability occurred as soon as the gas was removed from the proto-planetary disk, rather than at a late time. In this case, obviously, the giant planet instability cannot be the source of the Late Heavy Bombardment of the terrestrial planets, unlike postulated in the Nice model (Gomes et al., 2005; Levison et al., 2011). However, the very existence and the nature of the LHB is still debated (e.g. Hartmann et al., 2000), so this possibility cannot be excluded with absolute confidence.

In practice, though, the giant planets' eccentricities were damped by scattering planetary embryos and planetesimals during terrestrial accretion.  So in order to end up on their current orbits the giant planets' early orbits must have been {\it more} eccentric than their current ones, with $e_{J, S} \approx 0.07-0.10$; this configuration was called EEJS in Raymond et al. (2009) and Morishima et al. (2010). The problem is that no simulations of the giant planet instability (either in its ``early'' or ``late'' versions) that successfully reproduced the outer solar system ever succeeded in producing an orbit of Jupiter about twice more excited in eccentricity than the current one (Nesvorny, 2011; Nesvorny and Morbidelli, 2012). For this (and other) reasons, the EEJS configuration is not considered realistic in Raymond et al. (2009).

If one neglects these issues, the EEJS model can naturally produce a small Mars starting from a shallow surface density profile of the solid distribution ($x\sim 1.5$), because material is removed in the vicinity of the Mars region by the action of the $\nu_6$ secular resonance at 2.1 AU, which is initially super-strong in view of the larger eccentricity of Jupiter (Raymond et al., 2009). In this scenario, a lot of mass beyond $\sim$2.5 AU is also removed from the system because of the strong gravitational perturbation of Jupiter and Saturn.

However, for the goal of the current paper, which is to investigate whether a steep radial distribution of solids could explain simultaneously the small mass of Mars and the asteroid belt orbital properties, the EEJS set-up would not help. In fact, a steep surface density profile would have a small mass in the Mars region to start with, and the strong local depletion acted by the $\nu_6$ resonance would further reduce the   mass and in the end Mars would be too small. In terms of the asteroid belt, a more eccentric Jupiter would not substantially enhance the dynamical excitation of the asteroid belt, away from the $\nu_6$ resonance (Figure 11). Thus, we can conclude that the steep density profiles are unlikely to explain the structure of the inner solar system whatever the set-up of the giant planet orbits.

\subsection{An alternate model with a localized mass depletion (Izidoro et al., 2014)}

Our present paper was strongly motivated by the results of Izidoro et al. (2014), who modeled the formation of the terrestrial planets by assuming a local mass depletion between $\sim$1.1-1.3 and $\sim$2-2.1 AU in the original solid mass distribution (Jin et al., 2008). In this section we revisit this scenario discussing the success and limitations of that model.
 
Izidoro et al. (2014) claimed success in producing Mars-analogs and an excited and depleted asteroid belt in simulations considering initially Jupiter and Saturn as in their current orbits and a high mass depletion between 1.3 and 2 AU. One of the drawback in the results of Izidoro et al. (2014), however, is that simulations considering the giant planets in almost circular and coplanar orbits --as envisioned by models of the dynamical evolution of the solar system-- failed to produce Mars-mass planets around 1.5 AU (Morbidelli et al., 2012; Raymond et al., 2014; Raymond \& Morbidelli 2014). In Izidoro et al. (2014) the eccentric orbits of the giant planets, combined with an assumed severe initial local mass-depletion at around 1.5 AU (between 1.3 and 2 AU), are the key to produce Mars-analogs. This is because  the giant planets' gravitational perturbations  are stronger in this case (relative to the case where Jupiter and Saturn have almost circular and coplanar orbits) and help quickly removing a significant fraction of the $\sim$2 Earth masses  of solid material initially beyond 2 AU. This avoids that planets (Mars-analogs candidates) forming around 1.5 AU accrete too much mass. However, Jupiter and Saturn most likely  had  different orbits at the time the terrestrial planets were still forming (Section 4.3).

Similarly to the results of our shallow disks and other classical simulations (Chambers, 2001; Raymond et al., 2006; O'brien al., 2006), Izidoro et al.'s (2014) simulations also produced an excited and depleted asteroid belt. Nonetheless, qualitatively, the final distribution of bodies in this region only poorly matches the observed one (Izidoro et al., 2014). Another problem of the Izidoro et al. (2014) scenario is that is needed a localized mass depletion in solids around 1.5 AU probably too extreme compared to results of disk hydrodynamical simulations (see Jin et al., 2008). The scenario presented in this paper is essentially an updated version of that in Izidoro et al. (2014) using a self-consistent approach which invokes the radial drift and pile up of mass due to gas drag, leaving a depleted region beyond 1 AU.  Our current approach is also consistent with an envisioned global picture of the solar system evolution (Levison et al., 2011), whereas Izidoro et al. (2014) is not.

\begin{figure}
\centering
\includegraphics[scale=.8]{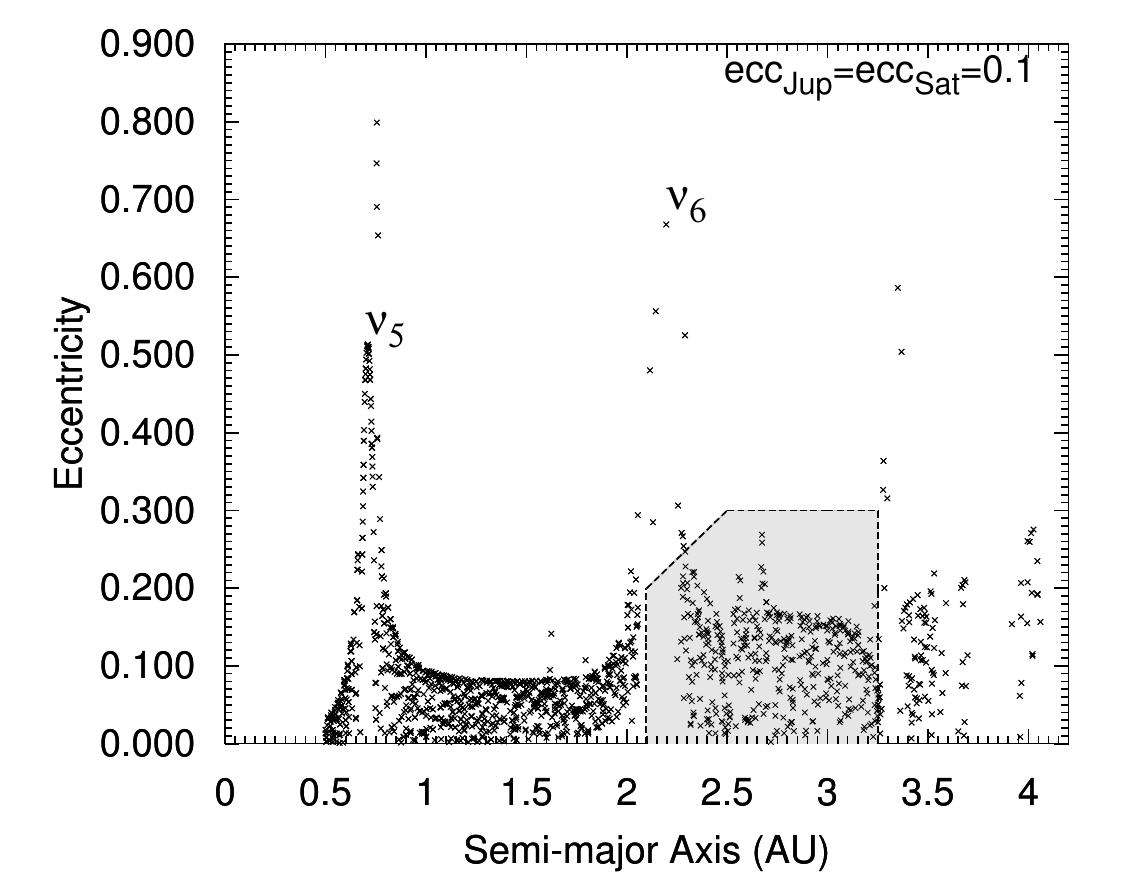}
\includegraphics[scale=.8]{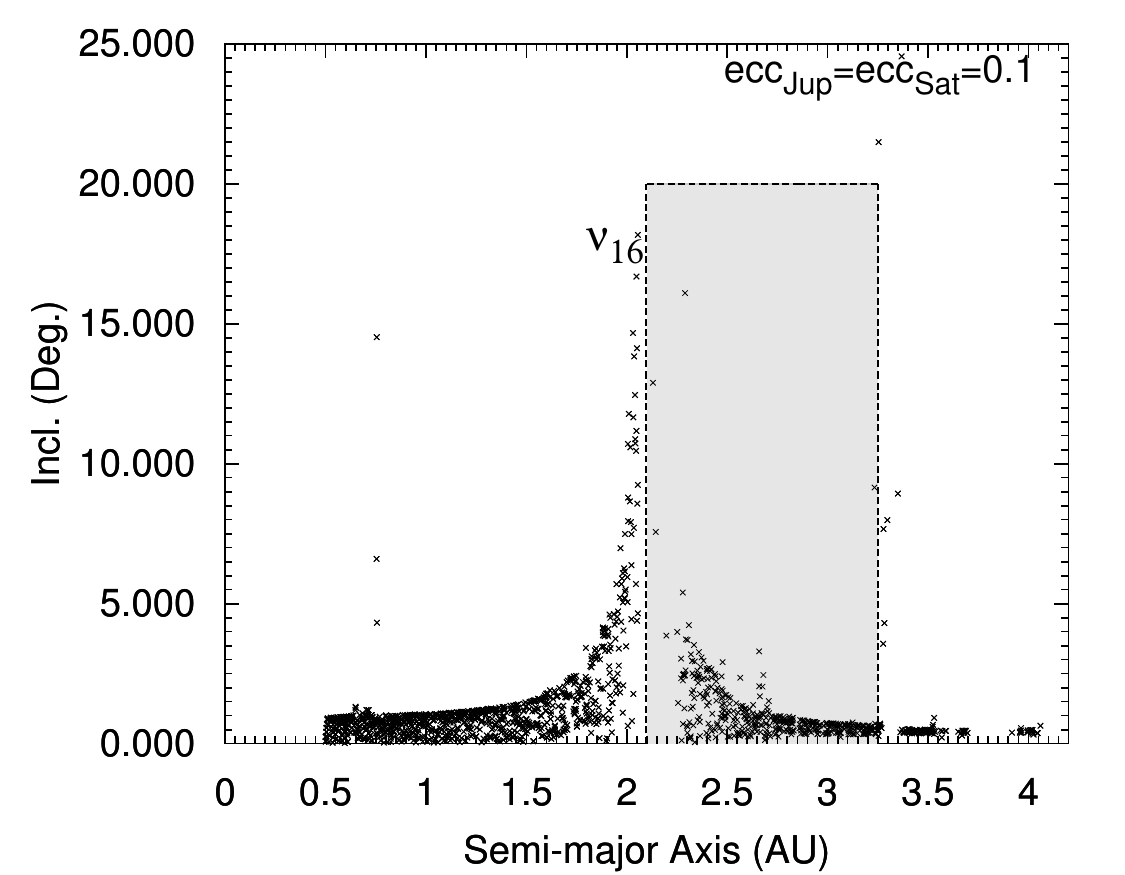}
\caption{ Snapshots at 10 Myr of a simulation considering 2000 test particles distributed between 0.5 and 4 AU. Jupiter and Saturn are initially as in their current orbits but with slightly higher orbital eccentricities. Secular resonance locations are also shown.}
\end{figure}

\section{Discussion}

The goal of this paper is to evaluate whether the inner Solar System could be reproduced by a disk of terrestrial building blocks with a power-law radial density profile.  We have assumed that planetary embryos and planetesimals are distributed from 0.7 to 4 AU and have tested four different surface density profiles: $\Sigma \propto r^{-x}$, where {\it x~}=~2.5, 3.5, 4.5 and 5.5, with a total of $\sim2.5 M_\oplus$ in solids.  We adopt the most recent version of the Nice model (Levison et al., 2011), which sets the giant planets' orbits to be resonant, nearly circular, and coplanar.

Our main result is that this scenario fails to match the properties of the inner Solar System.  A local mass deficit is needed to form a good Mars analog (e.g., Hansen 2009).  Yet a significant mass in planetary embryos in the asteroid belt is required in order to excite the asteroids' eccentricities and, more importantly, inclinations to their current values (e.g., Petit et al., 2001).  As expected from these arguments, only our steepest disk profiles ({\it x~}=~5.5) produced good Mars analogs, but those simulations yielded an under-excited asteroid belt.  Simulations with flatter disk profiles formed Mars analogs far more massive than the actual planet.  Those simulations  excited the asteroid belt to roughly the right amount but failed to adequately match observations because too many planetary embryos were stranded in the belt.

Our course, our simulations are limited.  For instance, we did not include the effect of non-accretionary collisions (see Leinhardt \& Stewart 2012; Stewart \& Leinhardt 2012).  But recent simulations by Chambers (2013) and Kokubo \& Genda (2010) have shown that the effects of fragmentation are minor and we do not expect this to qualitatively affect our results, i.e, to provide a solution to the small Mars problem. We also have considered a single initial total mass for the disk. However, more massive disks only tend to form more massive planets (e.g. Kokubo et al., 2006; Raymond et al., 2007). While higher mass disk may potentially alleviate at some level the cold asteroid belt issue produced in simulations with steeper disks,  a larger initial total amount of mass carried by protoplanetary embryos and planetesimals than we have considered would, in contrast, likely make it  more difficult to form Mars-analogs around 1.5 AU.

The mass pile-up in the inner solar system due to gas-drag would potentially  produce a very steep radial density profile, thus leading to a small Mars and a low mass asteroid belt. But our simulations imply that the asteroid belt would be too dynamically cold. This mechanism may be valid to explain the in-situ formation of the cold Kuiper belt (which has a very low mass and is indeed dynamically cold) but not the asteroid belt (which is dynamically hot). Interestingly, it is the asteroid belt, not the planetary system, that provides the most stringent constraint  against this pile-up model.

Given our result, the Grand-Tack model stands as the only current good explanation for both Mars and the asteroid belt. Of course, this does not mean that the Grand-Tack is correct. In fact, we acknowledge that there is a difficult synchronism in the growth/migration histories of Jupiter and Saturn  that is needed for the Grand-Tack to work (see Raymond \& Morbidelli 2014). Nevertheless there are as yet no satisfactory alternative models.

To summarize, we have shown that the simplest solution to the Mars problem -- a mass deficit beyond 1-1.5 AU -- creates a new problem by under-exciting the asteroid belt.  The formation of Mars and the current architecture of the asteroid belt are deeply connected constraints for models of Solar System formation (Izidoro et al., 2014).

\section*{Acknowledgements}

We are very grateful to the reviewer, Eiichiro Kokubo,  for his constructive comments and questions that helped us to improve an earlier version of this paper. A.~I, A.~M. and S.~R. thank the Agence Nationale pour la Recherche for support via grant ANR-13-BS05-0003-01 (project MOJO).  A.~I. also thanks partial financial support from CAPES Foundation (Grant:~18489-12-5). O.~C.~W thank support financial support from FAPESP (proc.~2011/08171-03) and CNPq. We thank the CRIMSON team for managing the mesocentre SIGAMM of the OCA, where these simulations were performed. A.I thank also  Helton Gaspar,  Rogerio Deienno, Ernesto Vieira Neto and Seth Jacobson for fruitful discussions.

\end{document}